%% file: ms.tex
\documentclass[fleqn,usenatbib,a4paper]{mnras}

\usepackage{amsmath,amssymb}
\usepackage{mathptmx}
\usepackage{txfonts}

\usepackage[T1]{fontenc}
\usepackage{ae,aecompl}
\usepackage{graphicx}
\usepackage{amssymb}
\usepackage{bm}
\include{journal}
\usepackage{color}

\newcommand{\teff}{\ensuremath{T_{\rm eff}}}
\newcommand{\beq}{\begin{equation}}
\newcommand{\eeq}{\end{equation}}
\newcommand{\intd}{{\rm d}}
\newcommand{\msun}{\ensuremath{M_\odot}}

\newcommand{\stefb}{\ensuremath{\sigma_{\rm SB}}}
\newcommand{\mdot}{\ensuremath{\dot{M}}}
\newcommand{\myr}{\ensuremath{\msun\,{\rm yr}^{-1}}}
\newcommand{\vcirc}{\ensuremath{v_{\rm circ}}}
\newcommand{\mpr}{\ensuremath{m_{\rm p}}}
\newcommand{\fa}{{\tt lowT\_fa05\_gn93}}
\newcommand{\freed}{{\tt lowT\_Freedman11}}
\newcommand{\ltwo}{L$_2$}
\newcommand{\vesc}{\ensuremath{v_{\rm esc}}}
\newcommand{\teq}{\ensuremath{T_{\rm eq}}}
\newcommand{\lsun}{\ensuremath{L_\odot}}
\newcommand{\ergs}{\ensuremath{{\rm erg}\,{\rm s}^{-1}}}
\newcommand{\tdust}{\ensuremath{T_{\rm dust}}}
\newcommand{\kb}{\ensuremath{k_{\rm B}}}
\newcommand{\kms}{\ensuremath{{\rm km\,s}^{-1}}}

\newcommand{\lthickmax}{\ensuremath{L_{\rm thick,max}}}
\newcommand{\lthin}{\ensuremath{L_{\rm thin}}}
\newcommand{\rc}{\ensuremath{r_{\rm c}}}
\newcommand{\tauzrc}{\ensuremath{\tau_{z, {\rm c}}}}
\newcommand{\tobs}{\ensuremath{t_{\rm obs}}}
\newcommand{\tscale}{\ensuremath{t_{\rm scale}}}

\title[Luminous Transients from Binaries]{Cool and Luminous Transients from Mass-Losing Binary Stars}

\author[Pejcha et al.]{ Ond\v{r}ej Pejcha,$^{1,3}$\thanks{pejcha@astro.princeton.edu}, Brian D.\ Metzger,$^2$ and Kengo Tomida$^1$
\vspace*{6pt}\\
$^1$ Department of Astrophysical Sciences, Princeton University, 4 Ivy Lane, Princeton, NJ 08540, USA\\
$^2$ Columbia Astrophysics Laboratory, Columbia University, New York, NY 10027, USA\\
$^3$ Hubble and Lyman Spitzer Jr.\ Fellow}

\pubyear{2015}

\begin{document}
\maketitle

%  proportional to the mass loss rate $\mdot$ and $\vesc$, 
%depend primarily on $\vesc$ and
% The predicted peak luminosities, timescales, and effective temperatures of mass-losing binaries are 

\begin{abstract}
We study transients produced by equatorial disk-like outflows from catastrophically mass-losing binary stars with an asymptotic velocity and energy deposition rate near the inner edge which are proportional to the binary escape velocity $\vesc$. As a test case, we present the first smoothed-particle radiation-hydrodynamics calculations of the mass loss from the outer Lagrange point with realistic equation of state and opacities. The resulting spiral stream becomes unbound for binary mass ratios $0.06 \lesssim q \lesssim 0.8$. For synchronous binaries with non-degenerate components, the spiral-stream arms merge at a radius of $\sim 10a$, where $a$ is the binary semi-major axis, and the accompanying shock thermalizes about $10\%$ of the kinetic power of the outflow. The mass-losing binary outflows produce luminosities reaching up to $\sim 10^6\,\lsun$ and effective temperatures spanning  $500 \lesssim \teff \lesssim 6000$\,K, which is compatible with many of the class of recently-discovered red transients such as V838~Mon and V1309~Sco. Dust readily forms in the outflow, potentially in a catastrophic global cooling transition. The appearance of the transient is viewing angle-dependent due to vastly different optical depths parallel and perpendicular to the binary plane. We predict a correlation between the peak luminosity and the outflow velocity, which is roughly obeyed by the known red transients. Outflows from mass-losing binaries can produce luminous ($10^5\,\lsun$) and cool ($\teff \lesssim 1500$\,K) transients lasting a year or longer, as has potentially been detected by {\em Spitzer} surveys of nearby galaxies.
\end{abstract}

\section{Introduction}

A fraction of binary stars evolve through at least one phase of dramatic angular momentum loss, accompanied by a drastic reduction of the semi-major axis and envelope mass ejection. The generally accepted framework for this stage of binary evolution is the ``common envelope'', when the drag forces on the binary orbiting in an non-corotating envelope lead to tightening of the orbit and ejection of the envelope \citep[e.g.][]{paczynski76,iben93,taam00,ivanova13a}. In some cases the binary may survive this phase with modified orbital parameters, becoming visible as a cataclysmic variable, X-ray binary or recycled pulsar \citep[e.g.][]{meyer79,bhattacharya91}. However, in other cases the two stars may merge completely into a single object. Perhaps the best known example is the progenitor of SN1987A, which is surrounded by a triple-ring structure, presumably ejected during such a merger \citep{chevalier89,hillebrandt89,podsiadlowski90,podsiadlowski92,morris07,morris09}.  Alternatively, the tightened orbit might lead to a delayed merger driven by gravitational wave emission or by an action of distant companions through the Kozai--Lidov process \citep[e.g.][]{thompson11,katz12,pejcha13}, ultimately involving the extreme physics of compact objects. This channel includes Type Ia supernovae important for distance measurements in cosmology \citep[e.g.][]{iben84,webbink84,weinberg13}, and neutron star and black hole mergers, which are promising sources for detecting gravitational waves \citep[e.g.][]{abadie10,metzger10}.

Despite its importance to many areas of astrophysics, observational tests of the common envelope framework have primarily been indirect and based on comparing observed post-common envelope binaries to theoretical predictions, often aiming to constrain the parameterized common envelope efficiency \citep{webbink84}. Recently, this dynamic and rapid phase of binary evolution has been associated with transients that have luminosities in between classical novae and supernovae, and red color \citep[e.g.][]{munari02,bond03,soker03,soker06,tylenda06,kulkarni07,thompson09,tylenda11,tylenda13,ivanova13b}. These transients, which include V838~Mon, M85~OT, M31~RV, and V4332~Sgr, are commonly named `luminous red novae', `intermediate luminosity optical transients', or `mergebursts' \citep{soker06}. Hereafter, we describe all of these events using the generic term `red transients' (RT), because some members of this class are as luminous as, or even dimmer than, classical novae.  The association between RT and binary stars has been strengthened by the pre-outburst photometry of V1309~Sco, which reveals an eclipsing contact binary with a rapidly decreasing orbital period \citep{tylenda11}. It has been proposed that RT optical emission results from the release of thermal and hydrogen recombination energy from a shell heated and launched by the binary interaction, similar to a shock passing through the hydrogen envelope of a Type II-Plateau supernova and leading to relatively cool effective temperatures \citep{ivanova13b}. RT occur every $\sim 2$\,years in Milky Way-like galaxies, a rate which is broadly consistent with binary population synthesis models \citep{kochanek14}. 

RT exhibit considerable diversity, with peak luminosities ranging from $10^3$--$10^4\,\lsun$ for V4332~Sgr and OGLE 2002-BLG-360 \citep{tylenda13} to $\sim 10^6\,\lsun$ for V838~Mon, or up to $10^8\,\lsun$ if objects such as SN2008S and NGC300 OT are included.  Explaining the latter two events also requires an unobserved shock breakout with a luminosity of $\sim 10^{10}\,\lsun$ \citep{kochanek11a}.  
Note also that the luminosity of the ``Rosetta stone'' event V1309~Sco was a factor of $\sim 10$ lower than V838~Mon and similar objects, and factor of $\sim 100$ lower than SN2008S and NGC300 OT. The outburst durations of RTs range from about $20$\,days for V4332~Sgr \citep{martini99} to roughly a year or longer for CK~Vul and OGLE 2002-BLG-360 \citep{kato03,tylenda13}. 

Fundamental differences also exist among the RT progenitors, when such identifications are possible. The progenitor of V1309~Sco was a binary containing an K-type primary \citep{tylenda11}, M31 LRN 2015 harbored a $2$--$4\,\msun$ evolved star \citep{dong15}, V838~Mon is known to have B3 main-sequence progenitor \citep{munari05,tylendaetal05b}, while the progenitor of SN2008S was a dust-obscured extreme asymptotic giant branch star \citep{prieto08,thompson09}.  The progenitor mass and luminosity appears to be correlated with the peak luminosity of the transient \citep{kochanek14}.

The light curves also display dramatic differences: some events show relatively smooth initial declines (SN2008S, NGC300 OT, V4332~Sgr), while others show multiple peaks (V838~Mon, OGLE 2002-BLG-360, CK~Vul), or evidence for sudden dust formation (M31 LRN 2015, SN2008S, NGC300 OT).  V1309~Sco exhibited a slow rise to maximum, which lasted $\sim 150$\,days -- a relatively unique feature among the RT, shared only by OGLE 2002-BLG-360.  However, a similar feature might have been missed in other objects. Finally, the RT also display distinct differences in their spectroscopic evolution.

The rich phenomenology of RT warrants a detailed analysis of the observable signatures of the dynamical phases of binary evolution.  By investigating the signatures of each process contributing to this evolution individually, we may hope to identify those discriminating characteristics which ultimately allow the observed diversity of RT to be mapped into individual astrophysical processes.  More ambitiously, by linking the remnants of the observed transients with known classes of objects, RT could one day serve as probes of the physics of the progenitors and their evolutionary pathways.  In this paper, we present the first steps in this direction.

One important channel to remove angular momentum from a binary star and shrink its orbit is through mass loss from the outer Lagrangian (\ltwo) point, a process which is efficient due to the long lever arm of the \ltwo\ point \citep[e.g.][]{pribulla98}. For \ltwo~mass loss to occur, both stars need to fill their Roche lobes and their common photosphere must expand sufficiently.  This condition can be achieved by an evolutionary change in the structure of one of the stars, which initiates a rapid, dynamical mass transfer \citep{webbink76}. The mass transfer rate will accelerate if there is positive feedback due to the adiabatic response of a donor if the latter possesses a deep convective envelope \citep{paczynski72,hjellming87,soberman97,ge10,ge15,woods11,pavlovskii15}. The secondary component cannot thermally accommodate the incoming material, swells up and fills its own Roche lobe.  Once the common surface reaches \ltwo\ as the two stars get closer by mass transfer, mass and angular momentum loss from \ltwo~will commence \citep[e.g.][]{webbink77}.  Alternatively, the binary can be brought into contact via a tidal instability \citep[e.g.][]{rasio95,eggleton01}, which also eventually results in dynamically unstable mass transfer and \ltwo\ mass loss \citep{tylenda11,stepien11}.

Matter leaves the \ltwo\ point in a spiral stream, which is torqued and accelerated by the time-changing gravitational force of the central binary. \citet{shu79} shows that for binary mass ratios $0.064 \lesssim q \lesssim 0.78$ the stream becomes unbound and escapes from the system.  For $q$ outside of this range, the stream remains bound and instead accumulates in a mass-loss `ring', which continues to interact with the binary.  The rest of this paper will focus predominantly on the unbound \ltwo\ outflows,  deferring a discussion of the observable signatures of the mass-loss ring to future work.  

Spiral streams leaving the binary spread radially with time, causing them to eventually collide and dissipate a fraction of their kinetic energy into heat and radiation by shocks. Although the merging of the spiral arms has long been expected \citep{shu79}, including the potential for shocks \citep{leibowitz81}, the importance of \ltwo\ mass loss events as transient electromagnetic sources has thus far been neglected. In this paper, we show that the \ltwo\ binary outflows can shine with high luminosities and low effective temperatures, similar to the observed properties of the RTs.

\ltwo\ mass loss may be of particular relevance for interpreting the light curve of V1309~Sco, which exhibits three unexpected features, as pointed out in \citet{pejcha14}.  First, the decrease in the orbital period of the contact binary directly observed in the $\sim 2000$-day light curve prior to the main outburst is rapidly accelerating; the second and higher period derivatives are incommensurably higher than the decay timescale $P/\dot{P}$. Second, the observed gradual changes in the phased binary light curve can only be understood if the binary is obscured anisotropically from one direction.  Finally, after the contact binary ceased to be visible, its gradual brightening to maximum lasted $\sim 150$\,days -- much longer than the dynamical timescale of the binary given by its orbital period of $P\approx 1.4$\,days.  \citet{pejcha14} argued that these three features are explained by \ltwo\ mass loss, which naturally ties together the period change, the obscuration of the binary star, and the brightening due to expansion of the pseudophotosphere in the outflow as the mass-loss rate runs away.  Although we defer a more thorough argument to future work, V1309~Sco strongly indicates that at least in some systems \ltwo\ mass loss runs away rather slowly, over many hundreds or thousands of binary dynamical timescales.  This justifies approximating the mass loss by a sequence of steady-state models.

Our results indicate that \ltwo\ mass loss leads to a universal configuration defined by a disk-like equatorial outflow, with energy deposited by colliding spiral streams at the inner edge. Both the energy deposition rate and the asymptotic velocity are proportional to the instantenous escape velocity from the binary. Similar outflow configurations are expected to occur also during some common envelope scenarios, when the released binary orbital energy leads to envelope heating and ejection \citep[e.g.][]{paczynski76,webbink84,iben93,taam89,livio90,sandquist98,taam00,morris06,ricker12,passy12,ivanova13a,macleod15}. Although the details of this process are not fully understood, the energy deposition rate and asymptotic ejection velocity likely scale with the orbital velocity of the binary. Our calculations are thus relevant for predicting the observable signatures of this considerably wider class of models, and we use \ltwo\ mass loss as an example of a relatively clean link between the binary properties and the energy deposition rate.

In this paper, we present the first radiation-hydrodynamics calculations of \ltwo\ mass loss, focusing on the observational appearance of these events. In Section~\ref{sec:setup}, we describe the numerical setup of our calculations. In Section~\ref{sec:struct}, we review the semi-analytic theory of the \ltwo\ spiral-stream outflow and analyze the numerical results. We also constrain the transport of the angular momentum from the binary to the \ltwo\ outflow, and estimate its gravitational and thermal stability. In Section~\ref{sec:an}, we study the radiation released from disk-like outflows with energy deposition at the inner edge and develop a semi-analytic model for luminosities and effective temperatures. In Section~\ref{sec:dust}, we study the dust formation in the outflow. In Section~\ref{sec:summary}, we summarize the observable signatures of \ltwo\ mass loss and compare them to known red transients. In Section~\ref{sec:disc}, we conclude by discussing extensions and other possible applications of the developed framework.

\section{Numerical setup}
\label{sec:setup}

\subsection{Smoothed Particle Hydrodynamics}

We study the structure of mass loss from the \ltwo\ point of binary stars using smoothed particle hydrodynamics (SPH). The advantages of SPH for this problem are the ability to handle anisotropic geometry, large density contrasts, and an accurate treatment of nearly-ballistic motion of the particles. We employ the conservative formulation of SPH with spatially varying smoothing lengths \citep{price07} and perform the calculations in a Cartesian inertial reference frame with the barycenter of the binary star as the origin. Our SPH particles move in a prescribed external time-varying gravitational field of the binary star in the center and we assume that the mutual gravitational interactions between the particles are negligible. The evolution of the velocity $\bm{v}_j$ of particle $j$ is
\begin{eqnarray}
\frac{\intd \bm{v_j}}{\intd t} &=& \sum_i  m_i\left(\frac{P_j}{\rho_j^2\Omega_j} + \frac{P_i}{\rho_i^2\Omega_i}\right)\nabla W_{ji} + \nonumber \\
&+& \sum_i m_i f_{ji}\Pi_{ji}\nabla W_{ji} +\bm{a}_{\rm binary},\label{eq:velocity}
\end{eqnarray}
where $m_i$, $\rho_i$, and $P_i$ are the mass, density, and pressure of particle $i$, respectively, $\Omega_i$ takes into account time-varying smoothing lengths, $W_{ji} = (W_j+W_i)/2$, where $W_i = W(r_{ji}, h_i)$ is the standard cubic smoothing kernel for smoothing length $h_i$ and the particle distance is $r_{ji} \equiv |\bm{r}_{ji}| \equiv |\bm{r}_j -\bm{r}_i|$, where $\bm{r} = (x, y, z)$ is the position vector. Following \citet{price07}, we simultaneously solve for $h_j$, $\rho_j$, and $\Omega_j$ by demanding that the mass within the smoothing volume stays constant. In each time step, we first predict an initial guess of $h_j$, and then refine the value using Newton-Raphson iteration. For rare cases when this procedure does not converge, which typically occurs for particles immediately after injection at \ltwo, we use bisection.

The second sum in Equation~(\ref{eq:velocity}) represents the artificial viscosity to resolve shocks, in particular we use the standard SPH viscosity of \citet{monaghan83},
\beq
\Pi_{ji} = \left\{  \begin{array}{cl}
      \frac{-\alpha_{\rm SPH}c_{S,ji}\mu_{ji} + \beta_{\rm SPH}\mu_{ji}^2}{\rho_{ji}}  & \bm{v}_{ji}\cdot \bm{r}_{ji} < 0, \\
      0 & \mbox{otherwise},
                   \end{array}
	  \right.
\eeq
where $\bm{v}_{ji} = \bm{v}_j - \bm{v}_i$ is the velocity difference of particles $j$ and $i$, $c_{S,ji} = (c_{S,j}+c_{S,i})/2$ is the mean adiabatic sound speed, $\rho_{ji} = (\rho_j + \rho_i)/2$, and
\beq
\mu_{ji} = \frac{h_{ji} \bm{v}_{ji}\cdot\bm{r}_{ji}}{r_{ji}^2 + \delta_{\rm SPH}h_{ji}^2},
\eeq
where $h_{ji} = (h_j + h_i)/2$ is the mean smoothing length, and $\alpha_{\rm SPH} =1.0$, $\beta_{\rm SPH}=2.0$, and $\delta_{\rm SPH}=0.01$ are coefficients. We find that reducing $\alpha_{\rm SPH}$ and $\beta_{\rm SPH}$ by a factor of $10$ leads to undesired particle interpenetration for our runs. The coefficient $f_{ji} = (f_j+f_i)/2$ is the \citet{balsara95} ``switch'',
\beq
f_i = \frac{|\nabla\cdot \bm{v}|}{|\nabla\cdot \bm{v}|+|\nabla\times\bm{v}|+10^{-4}c_{S,i}/h_i},
\label{eq:balsara}
\eeq
which limits the artificial viscosity in shear flows. We find that setting $f_{ji} \equiv 1$ does not noticeably change the results in this work, because there is no substantial shear present in our setup. We calculate the velocity derivatives for particle $j$ in Equation~(\ref{eq:balsara}) as
\begin{eqnarray}
\nabla\cdot\bm{v} &=& \sum_i \frac{m_i}{\rho_i} \bm{v}_{i}\cdot\nabla W_{ji},\\
\nabla\times\bm{v} &=& \sum_i \frac{m_i}{\rho_i} \bm{v}_{i}\times\nabla W_{ji}.
\end{eqnarray}
We experimented with other possible expressions, but did not find any noteworthy differences.

The last term in Equation~(\ref{eq:velocity}) is the gravitational acceleration due to the central binary
\beq
\bm{a}_{\rm binary} = -GM_1\frac{\bm{r}-\bm{r}_{*,1}}{|\bm{r}-\bm{r}_{*,1}|^3}-GM_2\frac{\bm{r}-\bm{r}_{*,2}}{|\bm{r}-\bm{r}_{*,2}|^3},
\eeq
where $M_1$, $M_2$, $\bm{r}_{*,1}$, and $\bm{r}_{*,2}$ are the masses and position vectors of the components of the binary with total mass $M$ and mass ratio $q = M_1/M_2$, $M_1 \le M_2$. The components of the binary move on a circular Keplerian orbit counter-clockwise in the $x-y$ plane with semi-major axis $a = |\bm{r}_{*,1}-\bm{r}_{*,2}|$ and orbital period $P$. The SPH particles do not affect the orbital properties of the binary. 

The internal energy $u_j$ of particle $j$ is evolved according to
\beq
\frac{\intd u_j}{\intd t} = \dot{u}_{{\rm gas},j} + \dot{u}_{{\rm visc},j} + \dot{u}_{{\rm diff},j} + \dot{u}_{{\rm cool},j} + \dot{u}_{{\rm heat},j},
\label{eq:u}
\eeq
where
\beq
\dot{u}_{{\rm gas},j} = \frac{P_j}{\rho_j^2\Omega_j}\sum_i m_i \bm{v}_{ji}\cdot\nabla W_j.
\eeq
 is the standard SPH term describing adiabatic expansion and contraction.  The viscous heating $\dot{u}_{{\rm visc},j}$ is calculated as
\beq
\dot{u}_{{\rm visc},j} = \frac{1}{2}\sum_i m_i f_{ji} \Pi_{ji} \bm{v}_{ji}\cdot\nabla W_{ji}.
\eeq

The remaining terms in Equation~(\ref{eq:u}) capture the radiative processes in the outflow. Although radiation transport has been previously included in SPH using implicit methods for flux-limited diffusion \citep[e.g.][]{whitehouse04,whitehouse05}, our goal in this work is to estimate, for the first time, the dependence of the radiative properties of the \ltwo\ mass loss on the parameters of the binary star and the mass loss rate. To these ends, we make several approximations in our treatment of radiative effects focused on making the radiative terms as local as possible and hence easily evaluated explicitly within the SPH formalism. Assuming that the radiation and the gas are in equilibrium, we split the radiative transport into three independent processes: flux-limited radiative diffusion $\dot{u}_{{\rm diff}}$, which only redistributes the energy between the particles predominantly in the radial direction, radiative cooling $\dot{u}_{\rm cool}$ in the vertical direction, and the irradiation by the central binary $\dot{u}_{\rm heat}$.  Accurate predictions of the emergent radiation should be done in the future using more sophisticated methods. 

The flux-limited radiative diffusion term $\dot{u}_{{\rm diff},j}$ \citep{bodenheimer90,cleary99,mayer07,forgan09} is evaluated as
\beq
\dot{u}_{{\rm diff},j} = \sum_i \frac{4m_i}{\rho_j\rho_i} \frac{k_j k_i}{k_j+k_i} (T_j-T_i)\frac{\nabla W_{ji}}{r_{ji}},
\label{eq:diff}
\eeq
where $T_i$ is the temperature of particle $i$, and $k_i$ is the thermal conductivity coefficient described in Appendix~\ref{app:diff}. Equation~(\ref{eq:diff}) models the diffusion of the radiation as a conduction process by interpolating between the optically-thick limit producing radiative diffusion and the optically-thin limit, when $\dot{u}_{{\rm diff},j}$ vanishes. Since the energy exchange between particles is pair-wise symmetric, equation~(\ref{eq:diff}) conserves energy and describes the redistribution of energy within the disk-like outflow. The energy losses due to radiative cooling, which occurs primarily perpendicular to the orbital plane, must be included separately.

\citet{stamatellos07} and \citet{forgan09} approximate radiative cooling and heating rates for SPH particles using local values of density, temperature, opacity, and gravitational potential. Their work was adapted to more anisotropic configurations by replacing the gravitational potential with the gas pressure scale height by \citet{lombardi15}. Neither of these prescriptions are immediately applicable to \ltwo\ outflows, because our SPH particles are not self-gravitating and their final trajectories do not deviate dramatically from the ballistic approximation (the outflow is not supported by the thermal pressure). Nonetheless, inspired by \citet{mayer07}, \citet{stamatellos07}, \citet{forgan09}, and \citet{lombardi15}, we assume that the radiative cooling occurs primarily in the direction perpendicular to the orbital plane and estimate the energy loss rate per particle as
\beq
\dot{u}_{{\rm cool},j} = -\frac{\stefb T_j^4}{\Sigma_{z,j}\tau_{z,j} + \kappa_j^{-1}},
\label{eq:cool}
\eeq
where $\Sigma_{z,j}$ and $\tau_{z,j}$ are surface density and optical depth from the position of particle $j$ out of and perpendicular to the orbital plane. This prescription interpolates between the optically-thin cooling rate of $\kappa_j \stefb T_j^4$ when $\tau_{z} \ll 1$, to a suppressed cooling rate as $\tau_z \gg 1$.  In the optically-thick limit, the energy is removed approximately on the diffusion timescale out of the outflow, $\dot{u}_{\rm cool} \approx u/t_{\rm diff}$ \citep{stamatellos07}. Although this suppression of radiative cooling is not very sophisticated, it has been successfully used in SPH simulations \citep{stamatellos07,forgan09,lombardi15}. However, a more realistic treatment of the radiation is definitely warranted. Note that the outflow can be optically-thick in the radial direction, yet optically-thin in the vertical direction, where most of the cooling occurs.  We set $\dot{u}_{{\rm cool},j} \equiv 0$ for $T<200$\,K to prevent particles cooling off the equation of state grid. 

In a similar manner to $\dot{u}_{{\rm cool},j}$, we include heating due to irradiation by the central binary as
\beq
\dot{u}_{{\rm heat},j} = \frac{L_*}{4\pi r_j^2}\frac{1}{\Sigma_{r,j}\tau_{r,j} + \kappa_j^{-1}},
\label{eq:heat}
\eeq
where $\Sigma_{r,j}$ and $\tau_{r,j}$ are surface density and optical depth in the radial direction.  The central binary is treated as a point source at the origin radiating isotropically with a bolometric luminosity of $L_* = 4\pi a^2 \stefb T_*^4$. The form of Equations~(\ref{eq:cool}--\ref{eq:heat}) guarantees that in the optically-thin limit the temperature of a stationary particle converges to the local equilibrium temperature given by
\beq
\teq = \left(\frac{L_*}{4\pi \stefb r^2}\right)^{1/4} = T_*\left(\frac{a}{r}\right)^{1/2}.
\label{eq:teq}
\eeq
The opacities $\kappa_j$ in Equations~(\ref{eq:cool}--\ref{eq:heat}) are technically Planckian means \citep{stamatellos07,forgan09}, but for the sake of simplicity we instead use the Rosseland means throughout this work.  Future refinements in this area should be accompanied by improvements in the radiation transport algorithm. More details of how we calculate the radiative cooling and irradiation heating are provided in Appendix~\ref{app:cool}. 

Because of the compact support of the smoothing kernel $W$, the summations in Equations~(\ref{eq:velocity}--\ref{eq:diff}) are performed only for a modest number of nearby particles. To find the nearest neighbors for our particles, we employ $k$-dimensional trees \citep{press07}. The SPH equations are integrated using the leapfrog method \citep{springel01}.  The global timestep $\Delta t$ is chosen to be the minimum over all particles of three timescales: (1) the Courant-Friedrichs-Lewy condition \citep{bate95},
\beq
\Delta t_{{\rm CFL}} = \frac{0.3h}{c_{S} + h|\nabla\cdot\bm{v}| + 1.2(\alpha_{\rm SPH}c_S + \beta_{\rm SPH}h|\nabla\cdot\bm{v}|)},
\eeq
where the last term of the denominator is included only for $\nabla\cdot\bm{v} < 0$, (2) the force condition
\beq
\Delta t_{\rm F} = 0.3\sqrt{\frac{h}{|\intd \bm{v}/\intd t|}},
\label{eq:tf}
\eeq
and (3) the condition that the internal energy should change by less than $10\%$ in one timestep,
\beq
\Delta t_{u} = 0.1\left|\frac{u}{\dot{u}}\right|.
\label{eq:tu}
\eeq
We have checked that our results are not sensitive to modest increases in the prefactors in Equation~(\ref{eq:tf}--\ref{eq:tu}), allowing us to adopt these less computationally expensive values.  

Since the studied problem does not necessitate the calculation of the gravitational forces between the particles, but requires frequent injection and removal of particles, querying microphysical tables, and evaluation of global quantities such as optical depths, we decided to write a new custom SPH code in the C/C++ language and parallelize using OpenMP. In addition to the $k$-dimensional tree routines of \citep{press07}, we use the GNU Scientific Library for root finding, sorting, and random number generation. The runtime of the application is dominated by the evaluation of the microphysics, and the global quantities such as optical depths, which is done efficiently in parallel. Construction of the particle tree and adding new particles are implemented serially, but this requires only very little computing power. Our code is explicit, which can lead to a very short, but not prohibitive, timestep in situations when the radiative diffusion dominates the energy equation. Typically, we require several tens or hundreds CPU hours to reach steady-state for optically-thin configurations, but some of our optically-thick runs took up to about $3000$ CPU hours on a 20-core machine.

\subsection{Luminosity and Emission Temperature}

We calculate the total radiative luminosity of the binary ejecta as a sum over all particles in the simulation,
\beq
L = \sum_j m_j|\dot{u}_{{\rm cool},j}|.
\eeq
The effective temperature of the emission, $\teff$, is estimated from the average of the effective temperatures of individual particles, assuming radiative equilibrium in a grey atmosphere and weighted by their radiative energy loss rate,
\beq
\teff^4 = \frac{1}{L} \sum_j m_j|\dot{u}_{{\rm cool},j}|\frac{T_j^4}{\tau_{z,j}+1}.
\eeq

We emphasize that our treatment of radiative losses and its descriptors $L$ and $\teff$ are only approximate.  Obtaining more accurate predictions of the luminosity and spectral energy distributions of mass losing binaries in future work will require using more sophisticated methods.

\subsection{Equation of state}

We use the equation of state (EOS) of \citet{tomida13,tomida15}, which assumes equilibrium abundances of ionized, neutral and molecular hydrogen including the rotational and vibrational degrees of freedom, and neutral, singly- and doubly-ionized helium. The composition is $X=0.7$, $Y=0.3$, and the ratio of ortho- to para-H$_2$ is assumed to be in thermodynamical equilibrium. The latent heats for all processes are consistently included. The EOS is given on a densely-spaced logarithmic grid spanning $-22 \le \log \rho \le 1.1$ and $0.2 \le \log T \le 6$. We modify the gas EOS by adding the contribution of radiation to the pressure and energy. However, we also need to modify the adiabatic sound speed $c_S$, which enters the calculation of artificial viscosity and time stepping conditions. Based on time-steady Euler equations, \citet[Appendix A]{pejcha12} expressed $c_S$ as a function of only derivatives of $P$ and $u$ with respect to either $\rho$ or $T$, while keeping the other quantities fixed. These thermodynamic derivatives can be easily evaluated on the gas EOS grid and we verify that the resulting gas $c_S$ is reproduced with better than $1\%$ accuracy. We then repeat the procedure with radiation contribution included in $P$ and $u$ to obtain the final $c_S$. Finally, we remap the EOS grid from the $\rho-T$ to $\rho-u$ plane to reduce the runtime computational overhead. At runtime, we obtain the thermodynamic quantities of interest by bi-log-linear interpolation.

\begin{figure}
\centering
\includegraphics[height=5.4cm]{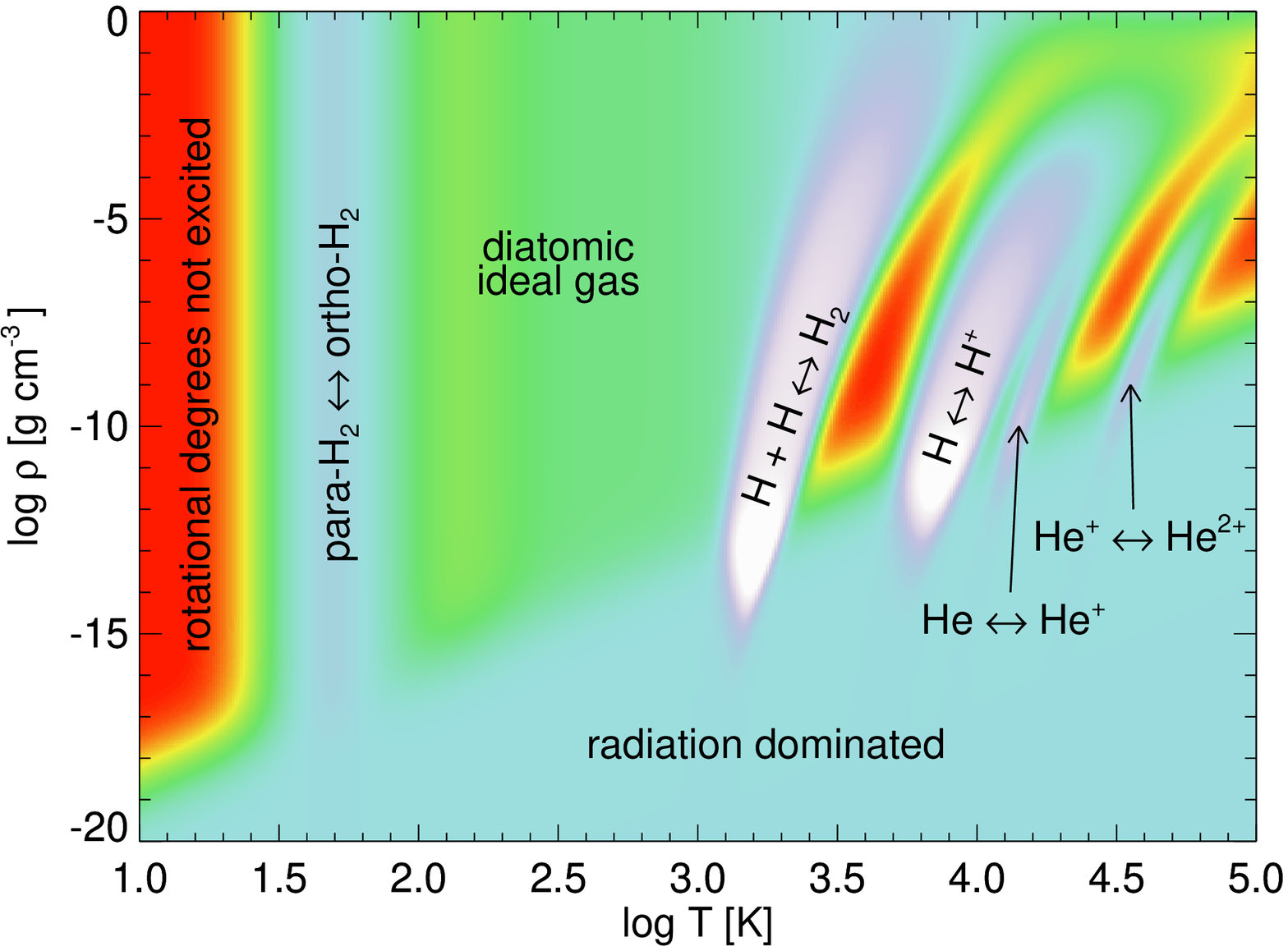}\includegraphics[height=5.4cm]{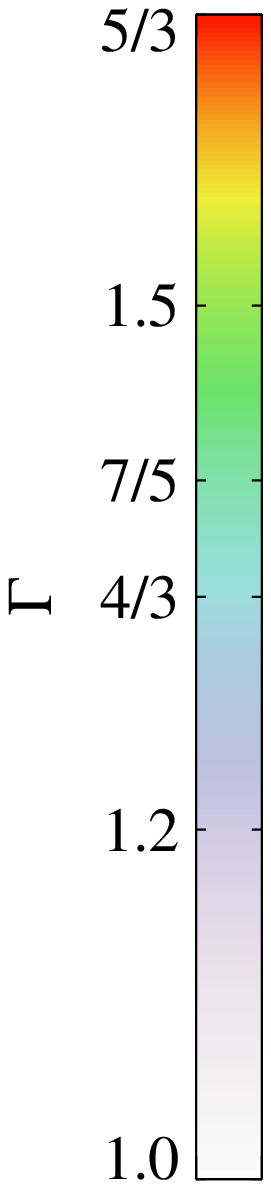}
\caption{Adiabatic index $\Gamma$ as a function $\rho$ and $T$ for our equation of state. Important state changes are labeled in the figure.}
\label{fig:eos}
\end{figure}

In Figure~\ref{fig:eos}, we summarize the basic features of the final EOS using the adiabatic index $\Gamma \equiv (\partial \ln P/\partial \ln\rho)_S = \rho c_S^2/P$. At low densities, the EOS is dominated by radiation with $\Gamma = 4/3$, while at higher densities the gas physics results in $1.0 \lesssim \Gamma \le 5/3$. Specifically, for $\log T \lesssim 1.5$, the EOS behaves as a mono-atomic gas with $\Gamma \approx 5/3$, because the temperature is not high enough to excite rotational degrees of H$_2$ molecules. For $1.5 \lesssim \log T \lesssim 2.0$, the conversion between para- and ortho-H$_2$ reduces $\Gamma$. For $2.0 \lesssim \log T \lesssim 3.0$, we find $\Gamma \approx 7/5$ corresponding to an ideal diatomic gas. For $\log T \gtrsim 3.0$, the latent heats associated with H$_2$ dissociation, and H and He ionization result in $\Gamma \approx 1$ for relatively large regions of the parameter space. The practical result is that if the energy input due to shock heating is comparable to the binding energy of H$_2$ ($2.26$\,eV per baryon) or the ionization energy of hydrogen ($13.6$\,eV), the temperatures will pile up between $1500$ and $3000$\,K or between about $8000$ and $12000$\,K, respectively. This is important for trying to analytically estimate the shock temperatures, as we discuss in Section~\ref{sec:an}.

\subsection{Opacities}

\begin{figure*}
\centering
\includegraphics[height=6cm]{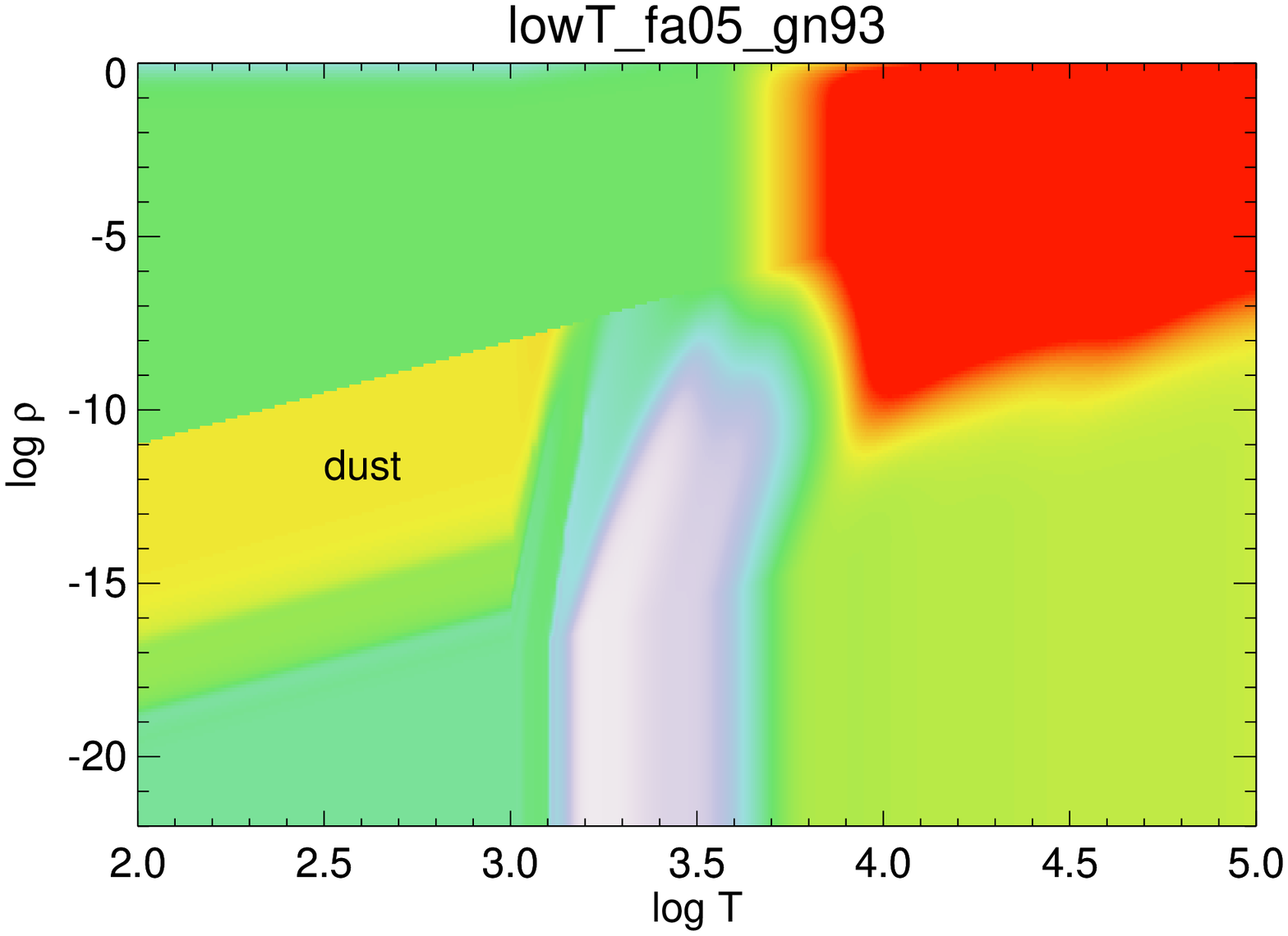}
\includegraphics[height=6cm]{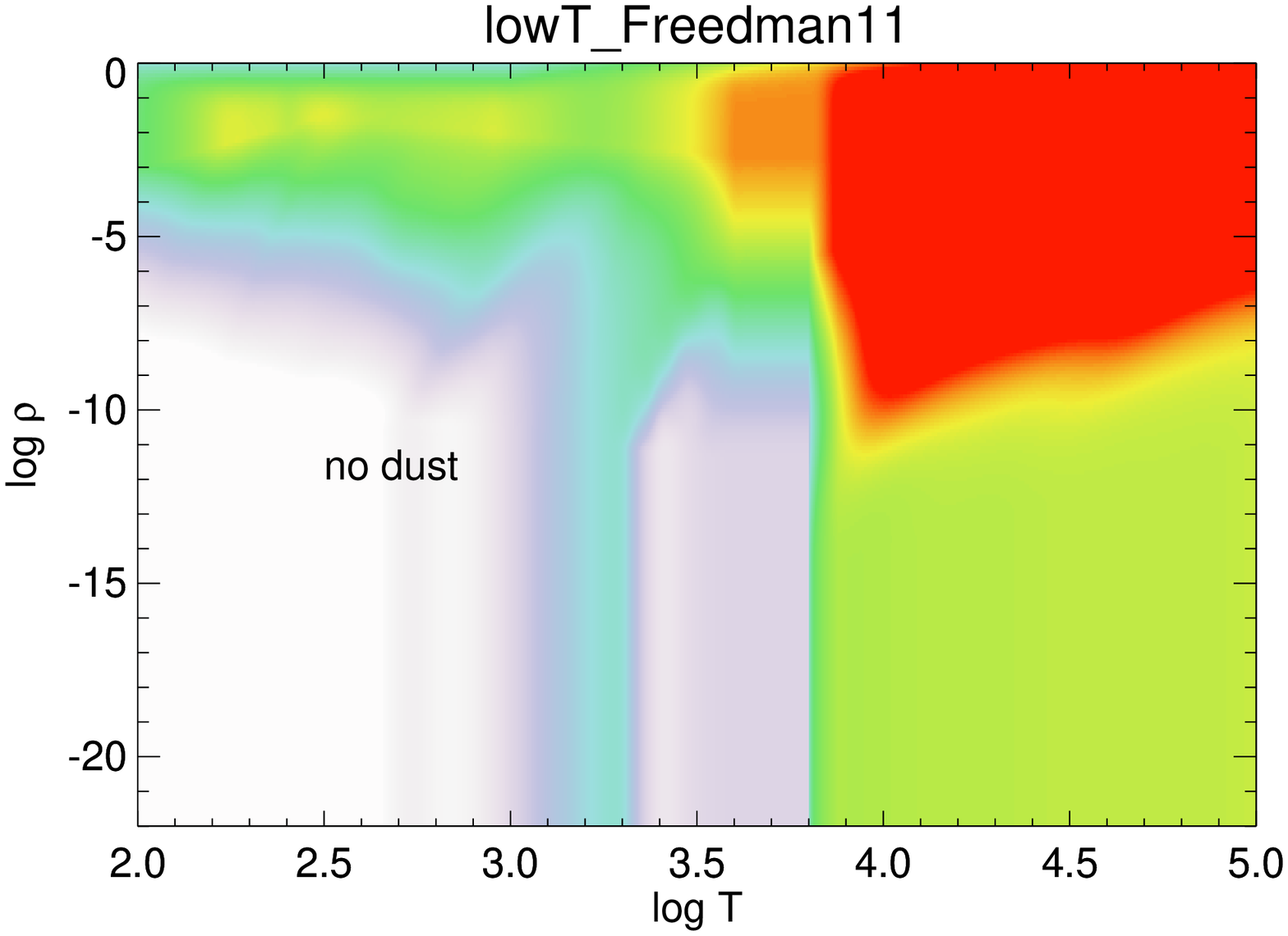}
\includegraphics[height=6cm]{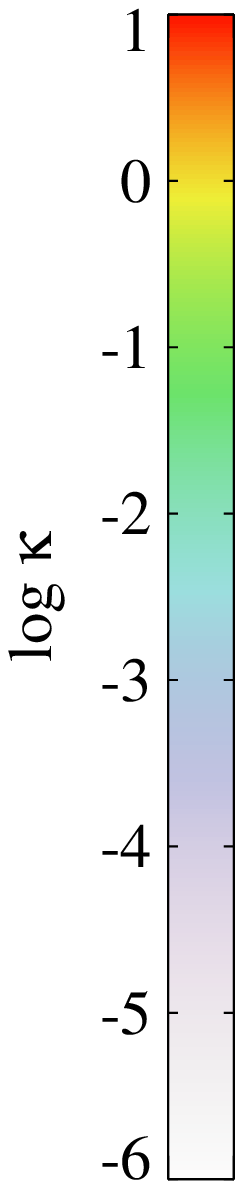}
\caption{Opacity tables as a function of temperature and density. Left panel shows low-temperature opacities from \citet{ferguson05} (option \fa\ in MESA), while the right panel is using opacities from \citet{freedman08} (option \freed). Outside of the boundaries of the input tables the opacities are set to the value at the boundary, which leads to stripes at low temperatures.}
\label{fig:kappa}
\end{figure*}

Opacities enter our calculation through the diffusion coefficients (Eqs.~[\ref{eq:diff}], [\ref{eq:app_ki}]), the estimates of radiative cooling (Eqs.~[\ref{eq:cool}], [\ref{eq:tau_zj}]), and irradiation by the central star (Eq.~[\ref{eq:heat}]). We employ the solar-metallicity $Z=0.0189$ opacity tables compiled in the stellar evolution code {\tt MESA} version 7503 \citep{paxton11,paxton13} by modifying the tutorial example {\tt sample\_kap.f} in the {\tt kap} module. Using this code we produce two opacity tables for the two possible choices of low-temperature opacity source by changing {\tt kappa\_low\_T\_prefix} option. The option \fa\ produces low-temperature opacities of \citet{ferguson05}, which include the effects of molecules and grains. The option \freed\ yields grain-free molecular opacities of \citet{freedman08} with updates to the molecular hydrogen opacity \citep{frommhold10} and ammonia opacity \citep{yurchenko11}. Both low-temperature opacity tables are blended with OPAL \citep{iglesias93,iglesias96} at high temperatures, as described in the MESA instrument papers \citep{paxton11,paxton13}. Opacities outside of the boundaries of the input tables are set to the values at the boundary. Although this potentially introduces bias in our calculations, it is probably rather small, because most of the radiation emission occurs at temperatures and densities covered by the tables.

The final opacity tables are visualized in Figure~\ref{fig:kappa}. For both tables, the opacity steeply rises for $\log T \gtrsim 3.7$ due to H$^-$ opacity. The opacities are very low for $3.1 \lesssim \log T \lesssim 3.6$ and $3.3 \lesssim \log T \lesssim 3.8$ for \fa\ and \freed, respectively. However, the tables significantly differ for $\log T \lesssim 3.1$, where the presence of dust grains in the \fa\ table results in relatively high opacities. Conversely, the grain-free table \freed\ exhibits very low opacities for low temperatures. Since the dust formation does not depend exclusively on the temperature, but also on other properties of the outflow (Section~\ref{sec:dust}), we employ both opacity tables and compare the results.

\subsection{Initial and boundary conditions}

Particles are injected around the \ltwo\ point with initial $y$ and $z$ coordinates drawn randomly from a Gaussian distribution of width $\varepsilon a$, where $\varepsilon \ll 1$ estimates the typical size of the \ltwo\ region and is defined following \citet{shu79} as
\beq
\varepsilon = \frac{c_T}{\vcirc} \approx 0.04 \left(\frac{T}{4000\,{\rm K}}\right)^{1/2} \left(\frac{M}{2\,\msun} \right)^{-1/2} \left(\frac{a}{0.1\,{\rm AU}} \right)^{1/2},
\label{eq:epsilon}
\eeq
where $c_T  = \sqrt{kT/\mpr}$ is the isothermal sound speed, $\mpr$ is the proton mass, and $\vcirc = \sqrt{GM/a}$ is the binary orbital velocity. The corresponding $x$ coordinate for each particle is fixed by requiring that it lies on a sphere centered on $M_1$ and intersecting the \ltwo\ point. We adopt a fiducial value of $\varepsilon=0.05$ (Eq.~[\ref{eq:epsilon}])  and investigate the sensitivity of our results in Appendix~\ref{app:technical}. The radial coordinate of the \ltwo\ point, $r_{{\rm L}_2} = \tilde{r}_{{\rm L}_2}a$, is determined by numerically solving the usual equation \citep[e.g.][]{shu79}. The initial velocities of the particles are set to the corotation velocity at the \ltwo\ point, $v_{{\rm L}_2} = 2\pi r_{{\rm L}_2}/P$. 

A particle is only injected into the full simulation if its purely ballistic motion keeps it exterior to \ltwo\ for a moderate initial `trial' period.  We experimented with different methods of injection, such as placing particles directly on the Roche surface, starting the stream slightly offset from \ltwo\, or imparting particles with an initial radial velocity kick of the order of $\varepsilon v_{{\rm L}_2}$. The results in each case were nearly identical, indicating that the specific injection procedure is not important.  For a given particle injection rate $\dot{N}$ and target mass loss rate $\mdot$, we determine the total mass of all active particles at each time step. If the actual number of particles is lower than the target, additional particles are added with constant mass $\mdot/\dot{N}$. Although the number of added particles is slightly different in each time step depending on the number of particles removed from the simulation, this procedure guarantees that we achieve the desired global time-averaged $\mdot$.  Typically, we choose $\dot{N} = 1000/P$ as our low-resolution calculation and explore the sensitivity of our results to $\dot{N}$ in Appendix~\ref{app:technical}.

Even if the ballistic trajectory of a particle would otherwise take it outside of \ltwo, pressure and viscous forces from neighboring particles might still drive it back inside.  Such particles would frenetically orbit the two point masses inside \ltwo, significantly reducing the global integration timestep and occasionally resulting in the violent ejection of the particle. We account for this clearly unphysical behavior by artificially removing any particles that reside closer to the coordinate origin or to $M_1$ than their injection position (usually such removals happen within a few timesteps after injection).  Typically, roughly one third of the initial particles survive to participate in the global dynamics.  The success fraction of particle injection increases if the injection point is moved slightly outside of \ltwo, although we find that such modest changes do not significantly alter the global outcome of the calculations. Our choice of an inflow inner boundary condition motivates our focus on binaries with $0.064 \lesssim q \lesssim 0.78$, which produce unbound outflows (Section~\ref{sec:ltwo}, \citep[Section~\ref{sec:ltwo} and][]{shu79}.

After particles are injected, we fix their temperatures artificially at the equilibrium temperature set by binary irradiation (eq.~\ref{eq:teq}) until the radial coordinate of the particle exceeds $1.4r_{{\rm L}_2}$, corresponding to a small fraction of the first spiral wrapping.  This approach removes transient effects that occur because some particles are injected supersonically with respect to others, in particular by mitigating shock heating within the stream very close to \ltwo.  Although such internal shocks are not obviously unphysical, our global results are not sensitive to the precise inner temperature evolution.  We have experimented with keeping the particles at a constant temperature (instead of $\teq$), or not masking this transient at all, finding no lasting influence on the structure of the stream beyond the first spiral wrapping.

\section{Structure of the \ltwo\ outflow}
\label{sec:struct}

\subsection{Mass loss from L$_2$}
\label{sec:ltwo}

Here, we summarize results on the semi-analytic treatment of mass loss from the outer Lagrange point \ltwo, which is based on the work of \citet{shu79}. Mass loss begins once the common surface of the synchronously rotating contact binary star comes into contact with \ltwo\ \citep{kuiper41}. Well inside of this surface, gas is supported in hydrostationary equilibrium by a balance between pressure and effective gravity (combination of gravitational and centrifugal accelerations). At \ltwo\ the confining force vanishes, allowing gas to transition to a hypersonic outflow. As argued by \citet{lubow75} for the L$_1$ point and \citet{shu79} for the \ltwo\ point, the gas-emitting region around \ltwo\ has a characterstic size $\varepsilon a$, with the transition to supersonic motion occuring on a similar spatial scale. The value of $\varepsilon$ depends only weakly on the parameters of the binary (Eq.~[\ref{eq:epsilon}]). Its small value indicates that, away from the immediate neighborhood of \ltwo, the thermal pressure of the star does not significantly affect the kinematics of the outflow (ballistic motion). 

\begin{figure}
\centering
\includegraphics[width=\columnwidth]{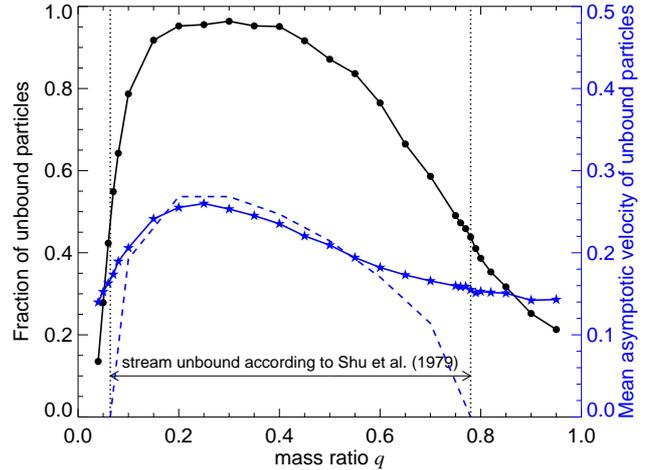}
\caption{Fraction of unbound particles ballistically ejected from the $\varepsilon$ vicinity of \ltwo\ (black solid line and circles) and their mean asymptotic velocity (blue stars) in the units of $\sqrt{2GM/a}$ as a function of the mass ratio of the binary. Dashed blue line shows asymptotic velocities from Table~2 of \citet{shu79}. The vertical dotted lines indicates the region, where \citet{shu79} found the stream escapes, $0.064 < q < 0.78$.}
\label{fig:leapfrog}
\end{figure}

As found by \citet{shu79}, only binaries with $0.064 \lesssim q \lesssim 0.78$ torque the gas enough to guarantee ejection to infinity.  We confirm this result by following the motion of a number of ballistic particles randomly injected near \ltwo\ from a neighborhood of size $\varepsilon a$.  The particles are assigned random velocity perturbations of the order of $\varepsilon v_{{\rm L}_2}$, where $v_{{\rm L}_2}$ is the corotation velocity at \ltwo.  Figure~\ref{fig:leapfrog} shows the fraction of particles that successfully escape to infinity.  We find that $\gtrsim 50$ per cent of the particles are unbound for $q$ within the limits defined by \citet{shu79}. Figure~\ref{fig:leapfrog}  also shows the mean asymptotic velocity $v_\infty$ of the unbound particles, which are typically between $0.15-0.25$ of the binary escape velocity, $\vesc = \sqrt{2GM/a}$. Our results for $v_\infty$ also closely match those of \citet{shu79}, except for values of $q$ with only a small fraction of unbound particles.  Since the ballistic problem is scale-free, these results are valid for all binaries. The relative insensitivity of $v_\infty$ to $q$ further suggests that measuring the \ltwo\ outflow velocity (from spectroscopy of a brightening merger event, for example) will constrain $\vesc$ and thus the total mass and semi-major axis of the binary.

In the rest of the paper we focus on binary mass ratios $0.064 \lesssim q \lesssim 0.78$ which produce unbound \ltwo\ outflows. For binaries with $q \lesssim 0.064$ or $q \gtrsim 0.78$, we confirm the finding of \citet{shu79} that gas piles up at a finite outer radius and forms a bound mass-loss ring instead of an unbound outflow.

\subsection{Density and temperature structure}
\label{sec:rho_t}

\begin{figure*}
\centering
\includegraphics[width=0.49\textwidth]{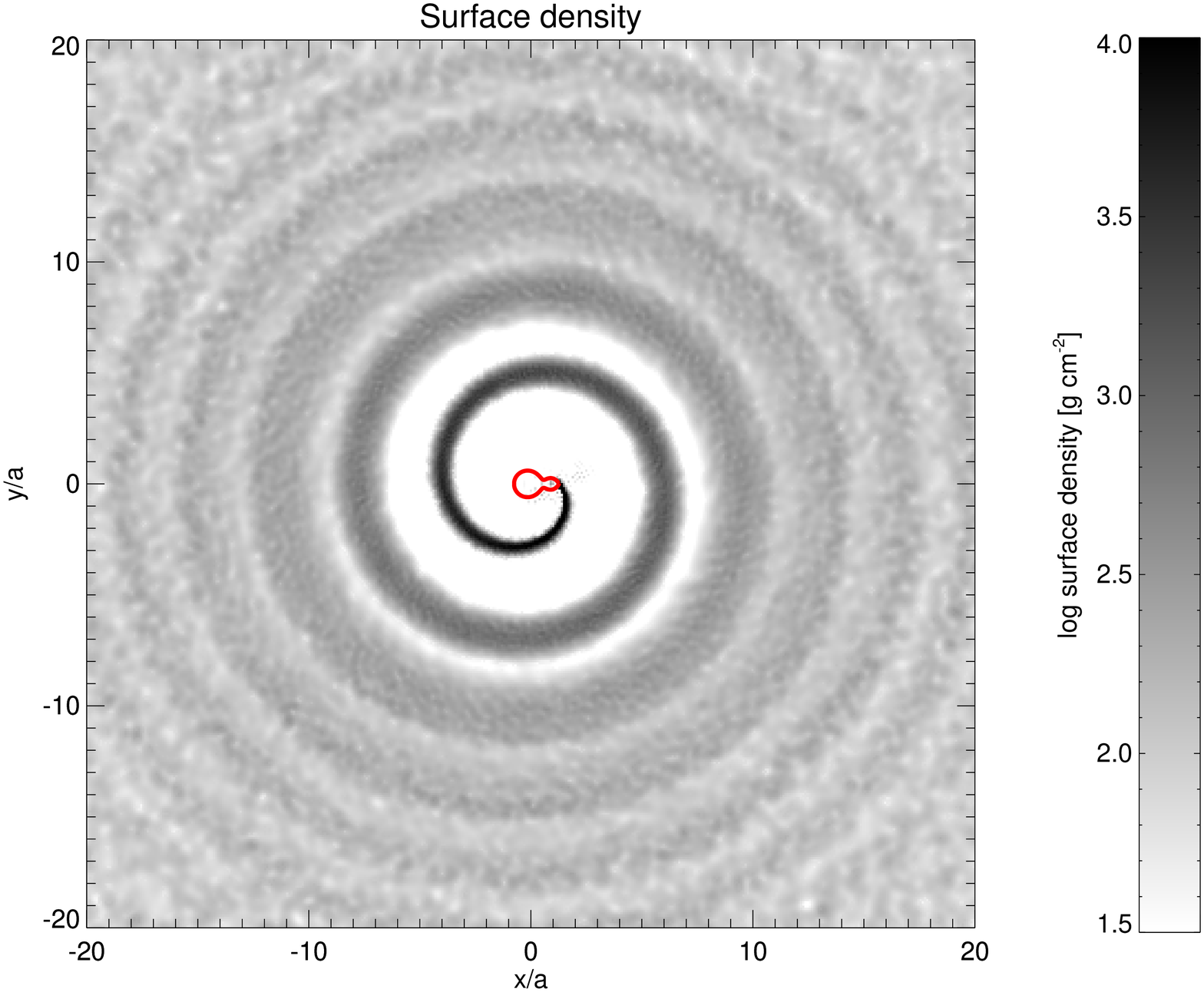}
\includegraphics[width=0.49\textwidth]{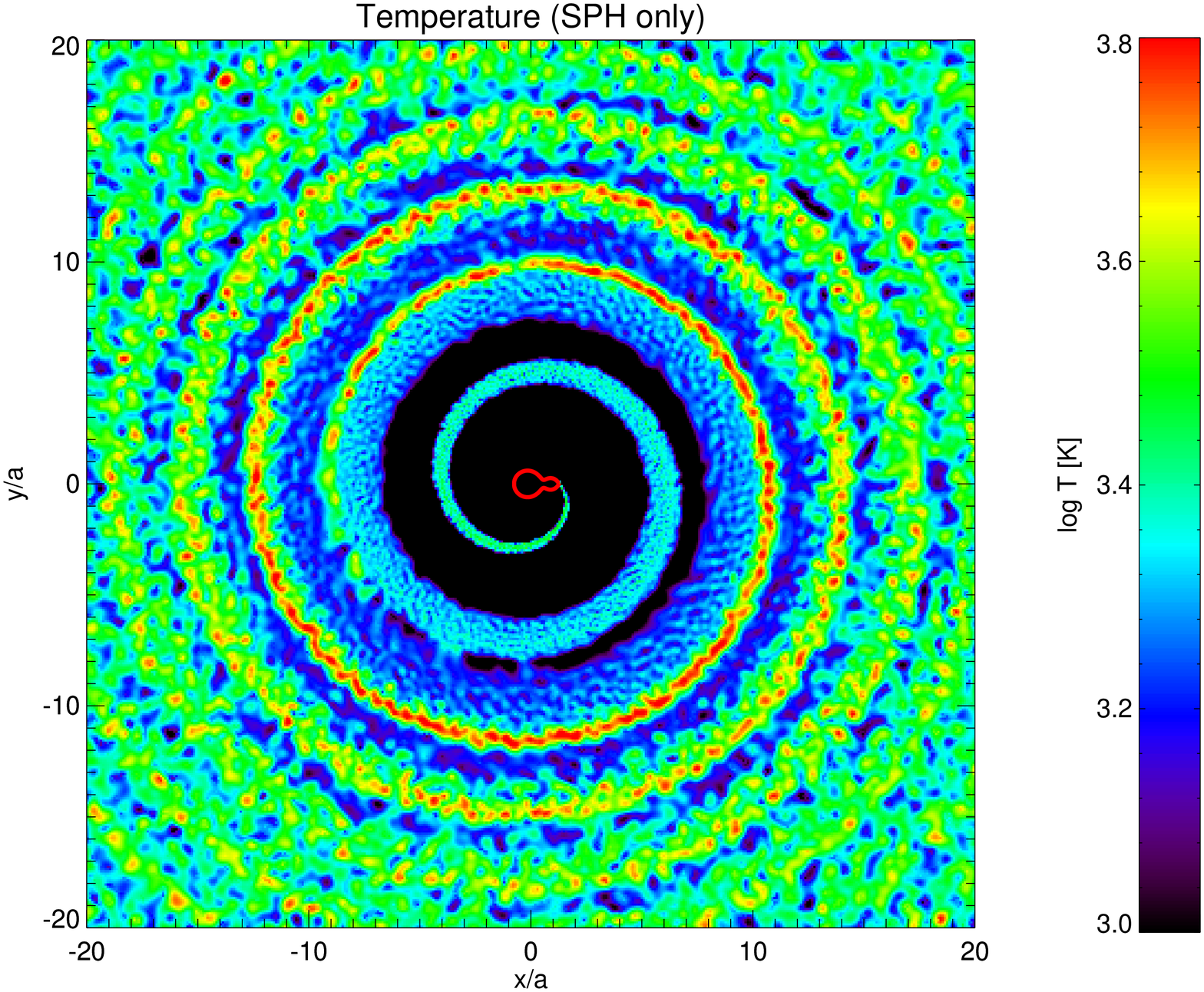}
\includegraphics[width=0.49\textwidth]{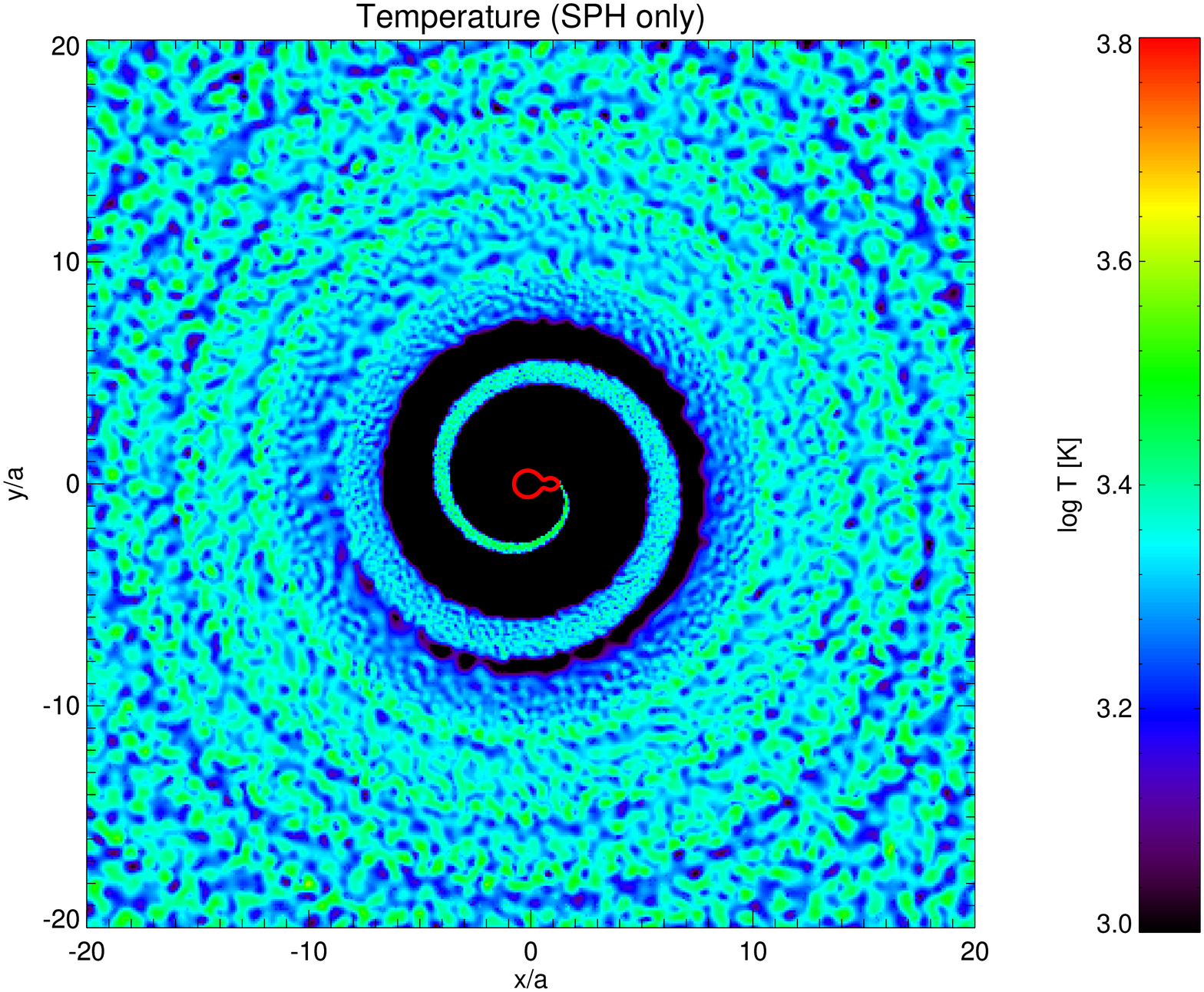}
\includegraphics[width=0.49\textwidth]{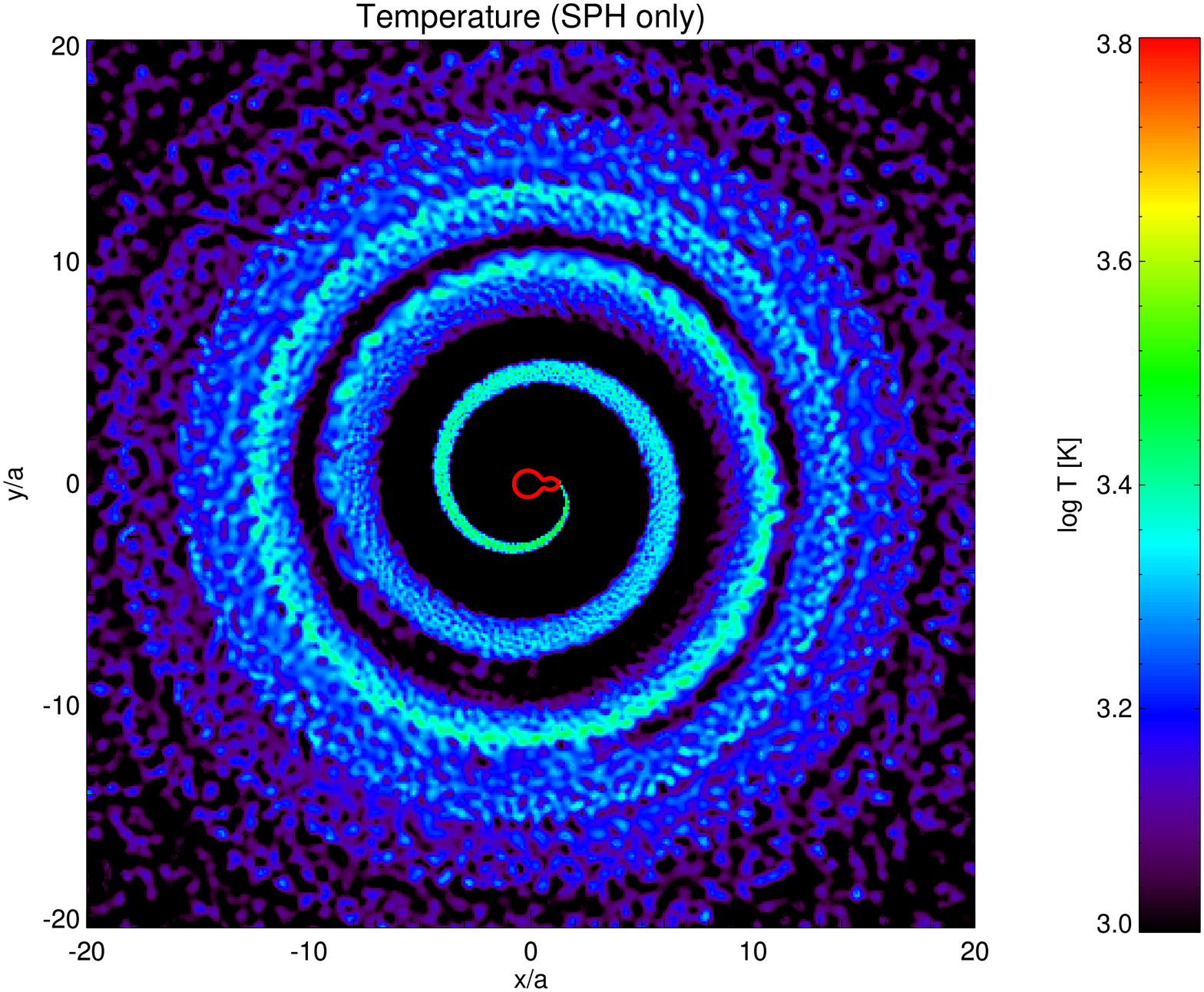}
\caption{Structure of the outflow from \ltwo\ for a binary with $M=1.725\,\msun$, $q=0.15$, $a=0.03$\,AU, and $\mdot=10^{-3}$\,\myr. The top left panel shows the surface density, while the remaining panels show the temperature for progressively more added physics. The potential surface containing \ltwo\ is indicated with a red line near the center.}
\label{fig:structure}
\end{figure*}

In Figure~\ref{fig:structure}, we show the typical density and temperature structure of the outflow from our radiation-hydrodynamics calculations. Initially, the stream of gas leaving \ltwo\ follows the spiral ballistic trajectory \citep{shu79}, while it expands and cools. The difference of gravitational forces across the stream contributes to the stream expansion. If the stream is optically-thin, then irradiation by the central binary compensates for the radiative cooling and adiabatic losses and the temperature converges to $T_{\rm eq}(r)$ (eq.~\ref{eq:teq}). For optically-thick streams, both irradiation and cooling are comparatively unimportant and adiabatic losses dominate. Although the stream expands significantly, the temperature drop is typically smaller than expected from a simple polytropic EOS due to the release of latent heat from H$_2$ formation. 

When the spiral wraps around the binary $\approx 1.5$ times and reaches the radial distance of $\rc \sim 8a$, the outer edge of the inner spiral encounters the inner edge of the previous wrapping of the spiral.\footnote{Note that $\rc$ depends on the initial spread of the stream and its temperature: if the stream was initially cold, $\varepsilon = 0$, no collision would occur between spiral wraps.  We can thus approximate $\rc/a \sim \varepsilon^{-1}$.} The associated velocity difference produces a shock, which heats part of the outflow. The thermodynamic properties at $r \gtrsim r_{\rm c}$ are set by the vertical optical depth at $r_{\rm c}$ and the relative importance of the individual terms in the energy equation. We illustrate this in Figure~\ref{fig:structure} by showing how the outflow temperature and density structure changes as the individual terms in Equation~(\ref{eq:u}) are switched on one by one. 

If the radiative diffusion and cooling are unimportant, the shock-heated material is advected outward forming a spiral (top right panel of Figure~\ref{fig:structure}). Typical peak temperatures are $\approx 10^4$\,K, depending on binary properties. If $\dot{u}_{\rm diff}$ instead dominates the energy evolution and radiation cooling is negligible, then the shock-dissipated energy is efficiently redistributed within the outflow and the spiral structure is erased (bottom left panel of Figure~\ref{fig:structure}). If the outflow is optically-thin in the vertical direction, then radiative cooling dominates, and the shocked gas cools almost immediately at the position where the two spiral wrappings first intersect (bottom right panel of Figure~\ref{fig:structure}). If the outflow is optically-thick, then the heated gas needs to be advected to larger radii before it can efficiently radiate and part of the thermal energy is then lost to adiabatic expansion.

\begin{figure}
\centering
\includegraphics[width=\columnwidth]{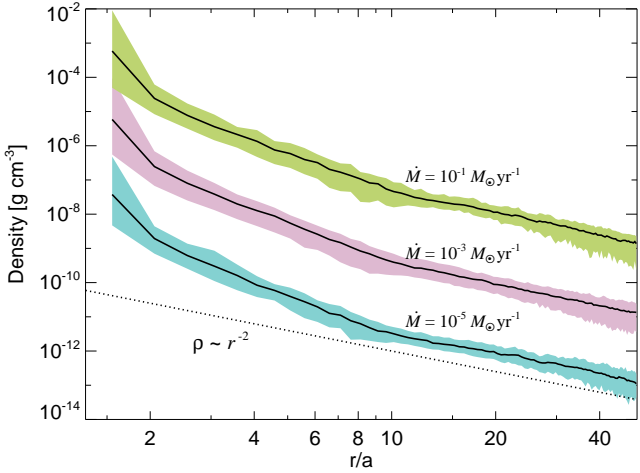}
\caption{Outflow density as a function of radius $r$ for the same binary as in Figure~\ref{fig:structure}, but for a range of mass loss rates $\mdot$.  Black lines show the median of the particle distribution, while colored bands show the $5-95\%$ percentile range.  For $r \gtrsim 10a$ the density profile obeys $\rho \propto r^{-2}$.}
\label{fig:rho}
\end{figure}

In Figure~\ref{fig:rho}, we show the density profiles as a function of the radial distance $r$ from the binary center. Initially, at small $r$ the density profile decreases steeply, $\rho \propto r^{-4}$. For $r \gtrsim \rc \sim 8a$, the density profile then flattens and behaves approximately as $\rho \propto r^{-2}$. This asymptotic behavior is useful in constructing a semi-analytic model in Section~\ref{sec:an}. There is no noticeable increase of density around $\rc$.  There are no remarkable differences in the density profiles for different binary configurations and mass-loss rates, except for its overall normalization.  

\begin{figure}
\centering
\includegraphics[width=\columnwidth]{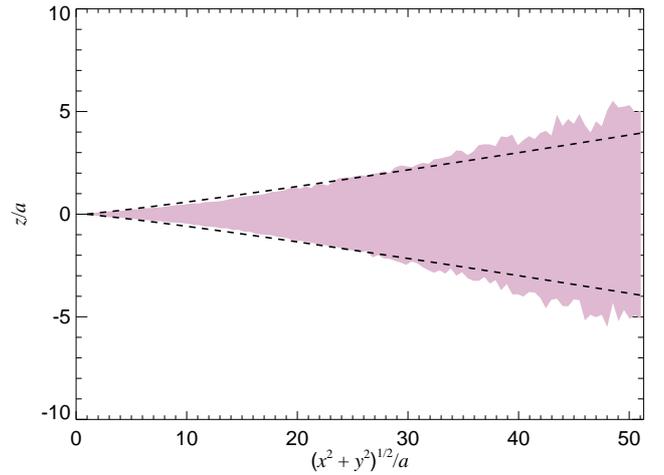}
\caption{Vertical distribution of particles as a function of the distance $\varrho = \sqrt{x^2+y^2}$ in the orbital plane, calculated for the same binary as in Figure~\ref{fig:structure}. The shaded area encompasses $5\%$ and $95\%$ of the particles at a given distance in the orbital plane.  Dashed lines show the expected growth of the thickness for free isothermal expansion in the vertical direction given by Eq.~(\ref{eq:zanal}) \citep{shu79}.}
\label{fig:rz}
\end{figure}

In Figure~\ref{fig:rz}, we show the vertical distribution of particles as a function of distance in the orbital plane $\varrho = \sqrt{x^2+y^2}$ and compare this to the analytic prediction of \citet[Eq.~61]{shu79}\footnote{We rewrite the result of \citet{shu79} by assuming that the particles move radially with a constant velocity. This is good approximation for $r \gtrsim \rc \sim 8a$, as we in Section~\ref{sec:energy_source}.},
\beq
z \propto (\varrho/a) \varepsilon \sqrt{2\pi\ln (\varrho/a)^2}.
\label{eq:zanal}
\eeq
Agreement between our numerical results and Equation~(\ref{eq:zanal}) is good if the constant prefactor is approximately unity.  Note that the thickness of the outflow  increases linearly with the relative size of the \ltwo\ region $\varepsilon$ (Eq.~[\ref{eq:epsilon}]) and faster than linearly with $\varrho$.  The outflow thickness is not found to depend sensitively on $\mdot$ or $q$.  For the fiducial $\varepsilon = 0.05$, the typical half-opening angles  at $\rc$ are between $5$ and $10$ degrees and the typical vertical thickness is  $\approx a$.

Finally, we investigated the effects of irradiation by the central binary and the alternative opacity tables on the results. Although irradiation heating primarily affects the temperature structure of optically-thin spiral streams, these changes are typically confined to the innermost spiral wrapping and do not result in qualitative changes in the spiral merging and associated energy release. Optically-thick outflows are found to reprocess up to about $10$ per cent of the binary luminosity.  We also compared our results for a series of calculations employing low-temperature opacities from \fa\ to those using \freed.  We find differences which are generally less than an order of magnitude, because most of the emission occurs under conditions of density and temperature where both tables provide similar results. The grain-free \freed\ opacities generally produce higher luminosities and slightly higher effective temperatures, because the lack of grains allows trapped radiation to escape earlier.  Since the conditions for dust formation are generally favorable in the outflow (Sec.~\ref{sec:dust}), we hereafter primarily employ the \fa\ opacity table, which includes condensates.

\subsection{Source of shock energy}
\label{sec:energy_source}

\begin{figure*}
\centering
\includegraphics[width=0.33\textwidth]{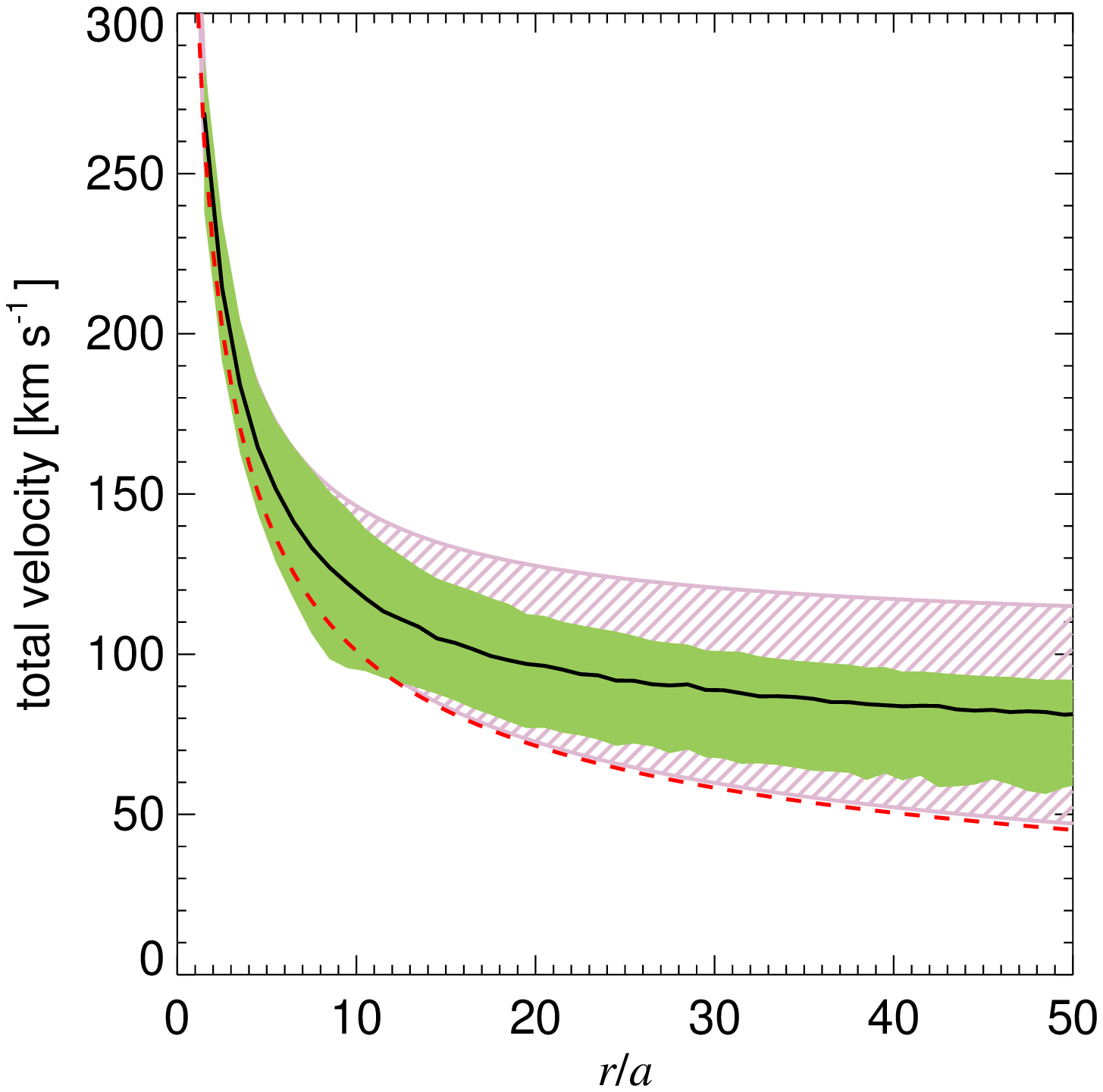}\includegraphics[width=0.33\textwidth]{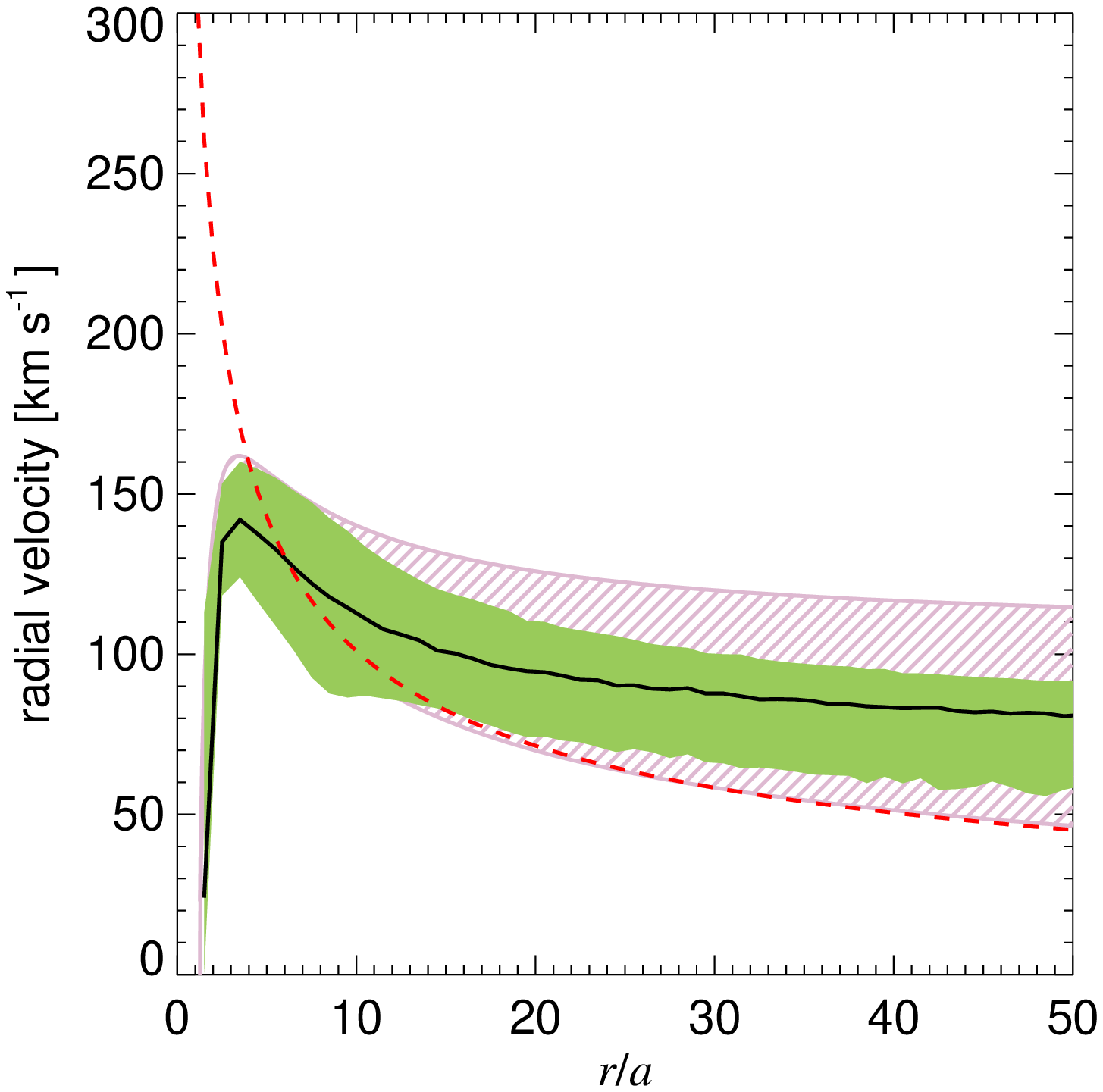}\includegraphics[width=0.33\textwidth]{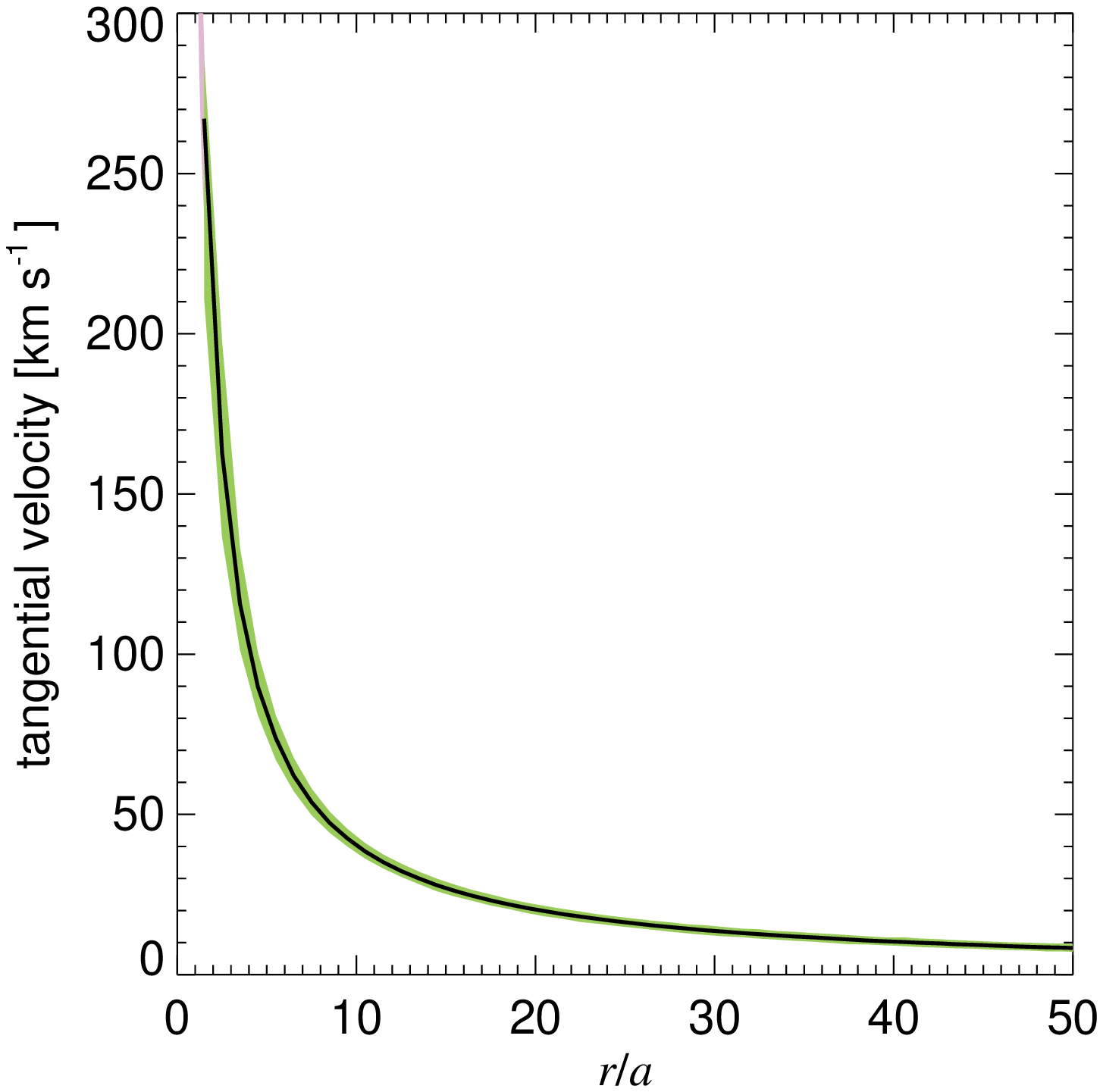}
\caption{Velocity distributions of particles as a function of radius, calculated for the same binary parameters as in Figure~\ref{fig:structure}.  We show the total velocity (left panel), radial component $|\bm{v}_{||}| = \bm{v}\cdot\bm{r}/r$ (middle panel) and the tangential component $|\bm{v}_\perp| = |\bm{v} - \bm{v}_{||}|$ (right panel).  Black lines show the median of the particle distribution, while green shading marks the $5-95\%$ percentile range of the distribution.  For comparison we show purely ballistic particle trajectories with similar initial conditions (pink hatched regions), and the local escape velocity $\sqrt{2GM/r}$ (red dashed line).}
\label{fig:velocity}
\end{figure*}

We now address the origin of the energy dissipated by the shock. Since the gas is moving highly supersonically, the kinetic energy dominates the total energy budget. In Figure~\ref{fig:velocity}, we show the velocity distribution of particles in one simulation snapshot and compare them to a range of purely ballistic trajectories with similar initial conditions. From the total velocity in the left panel, we see that the distribution of SPH particles, specifically the spread in velocity at a given radius, is initially nearly identical to ballistic particles. The particles are torqued and accelerated by the time-changing gravitational field of the binary, as evidenced by the growing disparity between the particle velocities and the local escape velocity $\sqrt{2GM/r}$. After particles are shocked at $\rc \sim 8a$, the velocity distribution becomes narrower, implying that faster moving particles transfer their energy to the slower moving particles. The velocity change $\Delta v$ is about $10$ to $20$ per cent of the escape velocity at $\rc\sim 8a$. In the middle and right panels of Figure~\ref{fig:velocity}, we show the radial and tangential components of the velocity. At $\rc$, the radial component dominates over tangential. There is little change in the tangential velocity distribution at $\rc$ and the distribution is narrow at all $r$. This implies that shock dissipation energy comes from the homogenization of the radial velocities of particles and that there is very little shear. This also explains why the simulation results do not depend on whether the \citet{balsara95} artificial viscosity ``switch'' is turned on or off.

\subsection{Instabilities in the outflow}
\label{sec:instabilities}

The outflow can be unstable to the gravitational formation of clumps, which could potentially violate our assumption of negligible gravitational interactions between the particles. The required condition is that the gravitational collapse time $t_{\rm grav} \sim (G\rho)^{-1/2}$ is shorter than the expansion timescale $t_{\rm exp} \sim r/v_\infty$. Assuming that $\mdot \sim r^2 \rho v_\infty$ (Fig.~\ref{fig:rho}), the stability condition is
\beq
\sqrt{\frac{G\mdot}{v_\infty^3}} \approx 0.02\ \left(\frac{\mdot}{0.1\,\myr}\right)^{1/2} \left(\frac{v_\infty}{100\,\kms}\right)^{-3/2} \ll 1.
\label{eq:grav_stability}
\eeq
The highest luminosities of the binary outflow investigated in Section~\ref{sec:summary} require not only high $\mdot$ but also high $\vesc$, and Equation~(\ref{eq:grav_stability}) implies that even these outflows are stable. This also justifies our assumption that the self-gravity among the particles is not important. The stability condition roughly holds at $\rc$ even if we replace the expansion timescale with the sound crossing timescale. 
%Eventually, as the outflow slows down by sweeping up circumstellar and/or interstellar medium and cools, the gravitational collapse into clumps becomes possible.

Alternatively, the outflow might form clumps by becoming thermally unstable. The condition for thermal instability roughly requires that the cooling rate increases along the thermodynamic trajectory of the gas parcel. This can happen during dust formation, when a small decrease of temperature leads to the increase of $\kappa$, which can yield higher cooling rate, more dust formation, and thus further increase in $\kappa$. In the context of multi-phase interstellar or intergalactic medium, conduction suppresses the formation of clumps smaller than the Field length
\beq
\lambda_{\rm F} = \sqrt{\frac{kT}{\rho\dot{u}_{\rm cool}}},
\label{eq:field}
\eeq
where $k$ is the conduction coefficient \citep[e.g.][]{field65,begelman90}. In this work, the energy in the outflow is redistributed by radiative diffusion, which is modeled as conduction in our SPH code (App.~\ref{app:diff}). The binary outflow becomes thermally unstable if $\lambda_{\rm F} \ll r$ and the clumps fit within a causally connected region. We evaluated $\lambda_{\rm F}$ in several runs covering various values of $\mdot$ and $\vesc$ and find that unless the radiative diffusion dominates the evolution, parts of the outflow are thermally unstable with $\lambda_{\rm F}/r$ between $0.01$ and $0.1$. Our simulations do not have sufficient resolution to demonstrate whether the clumping actually occurs or to study the properties of the clumps, but this topic deserves further attention. Clumping of the outflow might be responsible for the small quasi-periodic variations in the rising light curve of V1309~Sco \citep{pejcha14}.

\subsection{Backreaction on the binary orbit}
\label{sec:backreaction}

Gas leaving the binary carries away the angular momentum of the \ltwo\ point, $\dot{J} = 2\pi \mdot  (\tilde{r}_{{\rm L}_2}a)^2/P$. If the orbit remains circular, the orbital period of the binary shrinks as
\beq
\frac{\dot{P}}{P} = \mathcal{A} \frac{\mdot}{M},
\eeq
where $\mathcal{A}$ strongly depends on the mass ratio of the binary and on how does $q$ change throughout the mass loss \citep[e.g.][]{pribulla98}.

The gas leaves to infinity relatively ``cleanly'' in the sense that it interacts with the central binary only through the time-changing gravitational field and not directly through collisions and shocks around the stars. To assess the magnitude any potential additional angular momentum loss due to gravitational torquing, we calculate the total angular momentum of the outflow, $\bm{J} = \sum_j m_j \bm{r}_j \times \bm{v}_j$, and compare its value to the sum of angular momenta of particles at the time of their injection $\bm{J}_{\rm init}$. We find that $|\bm{J}|$ very soon asymptotes to values about $10$ to $18\%$ higher than $|\bm{J}_{\rm init}|$, depending on $q$. This is a relatively small correction to $\mathcal{A}$, especially considering the much greater influence of uncertainties in $q$ on $\mathcal{A}$ \citep{pribulla98}. For binaries forming a mass-loss ring ($q \lesssim 0.064$ or $q \gtrsim 0.78$), the angular momentum extraction might be much more efficient.

\section{Radiative properties  and Semi-Analytic Model}
\label{sec:an}

\subsection{Results from simulations}

\begin{figure*}
\centering
\includegraphics[width=0.49\textwidth]{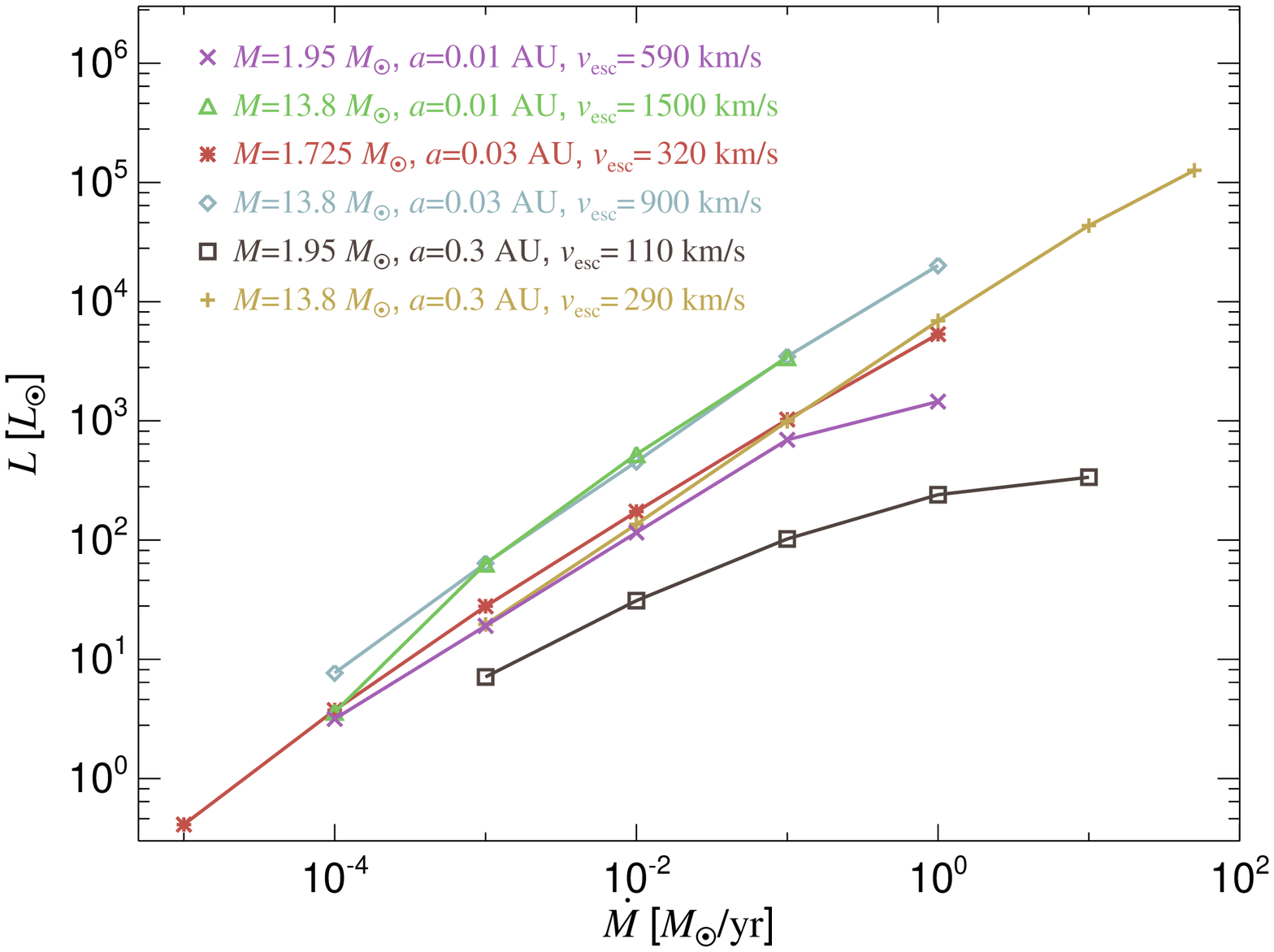}
\includegraphics[width=0.49\textwidth]{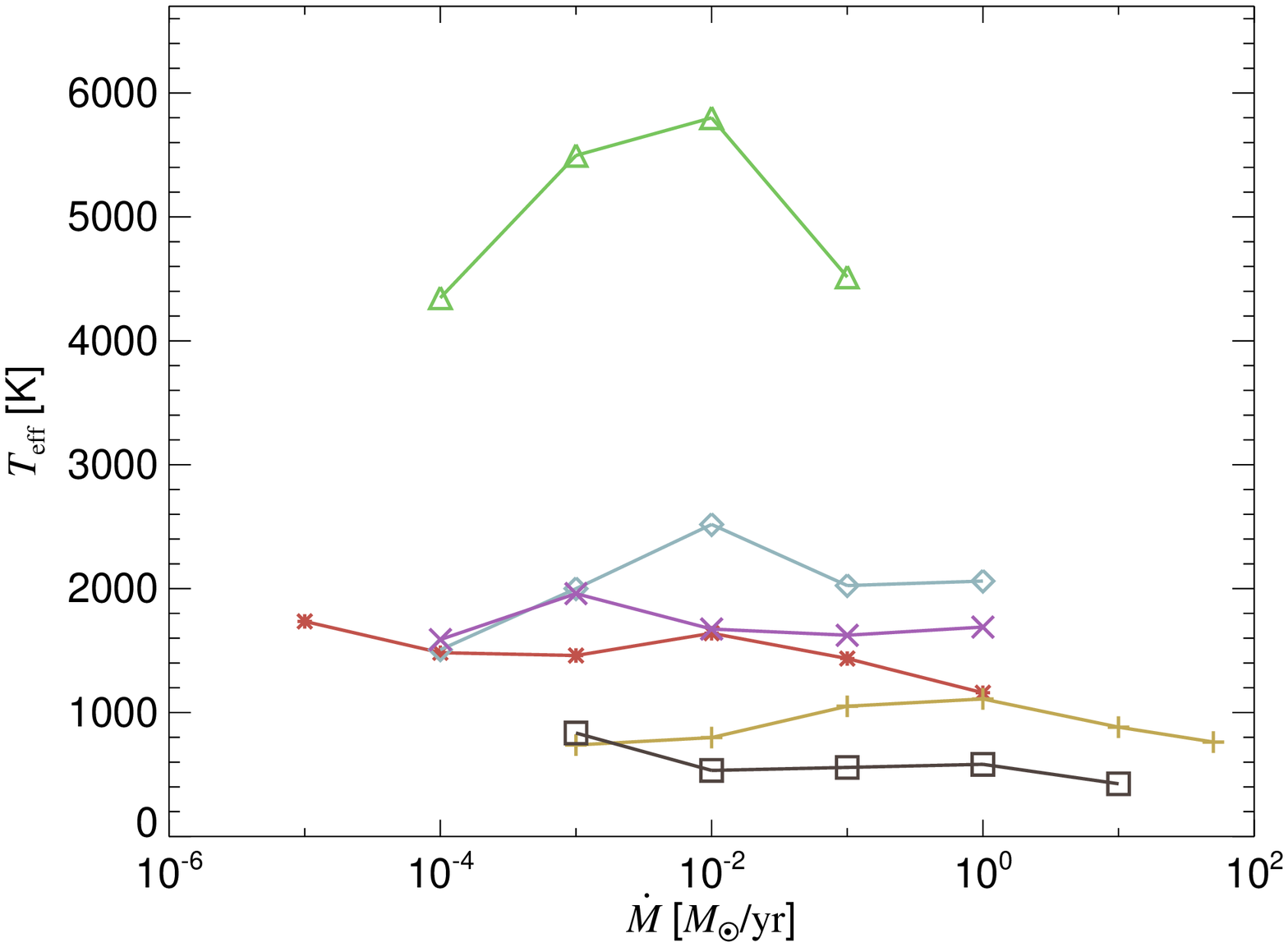}
\caption{Asymptotic radiative cooling luminosity (left panel) and the effective temperature (right panel) as a function of (temporally fixed) mass loss rate $\mdot$ for simulations which have achieved a steady-state outflow.  Different binary parameters ($M, a, \vesc)$ are denoted with symbols as described in the legend.  Irradiation by the central star was not included so as not to dominate the shock-powered emission at low $\mdot$.  In some cases the time needed to reach a steady state exceeds the time over which the binary can lose mass at the rate $\dot{M}$; we take this constraint into account when discussing applications of our results to red transients in Section~\ref{sec:summary}.}
\label{fig:lum}
\end{figure*}

In Figure~\ref{fig:lum}, we show the asymptotic luminosities and effective temperatures obtained by simulating binaries with constant $\mdot$ long enough for the outflow to establish a steady-state. The asymptotic luminosity is approximately linearly proportional to $\mdot$, while binaries with higher escape velocities are also more luminous.  For the range of parameters considered in Figure~\ref{fig:lum}, luminosities reach up to $10^6\,\lsun$. Note that the steady-state assumption may not be fully appropriate to describe some of the red transients, a point we discuss further in Section~\ref{sec:summary}.

The right panel of Figure~\ref{fig:lum} shows the  effective temperature of the emission. For the range of binary properties shown, $500 \lesssim \teff \lesssim 6000$\,K.  Although $\teff$ depends only very weakly on the mass-loss rate, a clear trend emerges in that binaries with higher escape velocities $\vesc$ produce higher temperature emission. This agrees with the physical picture presented in Section~\ref{sec:energy_source} that the shock dissipates a fixed fraction of the outflow kinetic energy $\propto \vesc^{2}$ (Section~\ref{sec:ltwo}).  The luminosities and effective temperatures derived from our simulations are similar to those of the observed red transients (Sec.~\ref{sec:summary}).

%\section{Dependence on the binary parameters}
%\label{sec:bin_pars}

%In Sections~\ref{sec:struct} and \ref{sec:an} we showed that the luminosity of the outflow depends primarily on the mass-loss rate and the escape velocity of the binary. Here, we explore the dependency on other binary parameters such as the mass ratio and the total mass of the binary.

\begin{figure}
\centering
\includegraphics[width=\columnwidth]{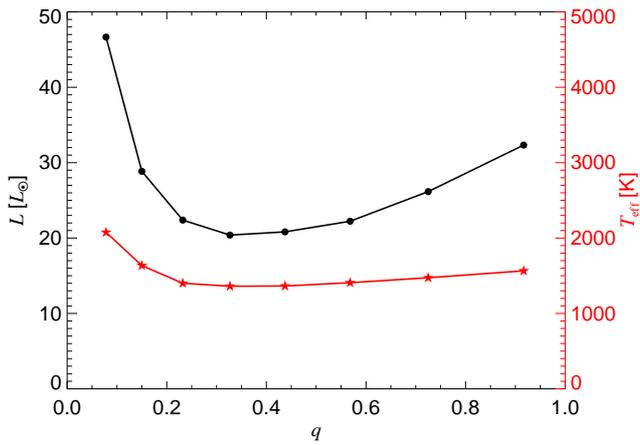}
\caption{Luminosity $L$ (black circles) and effective temperature $\teff$ (red stars) as a function of mass ratio $q$ for the binary from Figure~\ref{fig:structure}. The outflow is radiatively cooling in the optically-thin regime in all cases.}
\label{fig:sensitivity_q}
\end{figure}

In Figure~\ref{fig:sensitivity_q}, we show the asymptotic luminosity and effective temperature as a function of mass ratio, when all other parameters are held fixed. We see that both $L$ and $\teff$ exhibit minimum at $q \sim 0.3$, and increase on either side. This behavior can be understood by analyzing the motions of ballistic particles (Fig.~\ref{fig:leapfrog}), which show that the fraction of particles escaping to infinity decreases when $q$ is either large or small. As the efficiency of particles escaping to infinity decreases, the collisions become more efficient, resulting in a higher luminosity.  Since the luminosity is released at roughly the same distance from the binary, $\teff$ follows the same trend as $L$. In all cases, the luminosity varies by less than a factor of $2$ and hence the dependence on $q$ can be neglected in analytic estimates.

\begin{figure}
\centering
\includegraphics[width=\columnwidth]{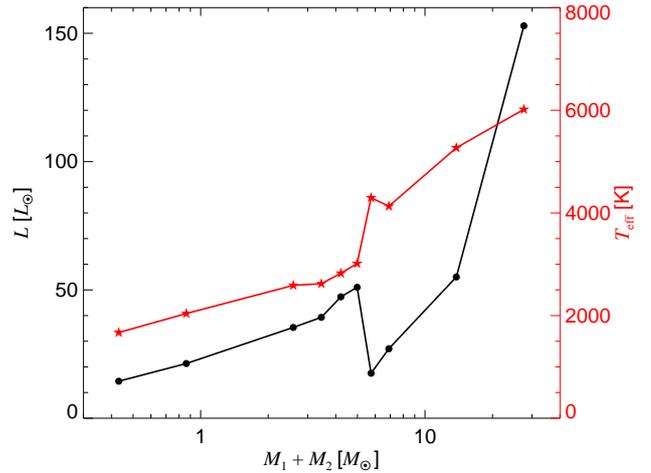}
\caption{Dependence of the luminosity (black circles) and effective temperature (red stars) on the total binary mass $M$ for fixed $a=0.01$\,AU, $q=0.15$, and $\mdot = 10^{-3}\,\myr$. The jump in curve is caused by the transition between diffusion-dominated and dust-dominated outflows, as explained later.}
\label{fig:sensitivity_m}
\end{figure}

In Figure~\ref{fig:sensitivity_m}, we show the dependence of the luminosity and effective temperature on the total mass of a binary with $a$ and $q$ held fixed. As expected, $L$ and $\teff$ increase with $M$, because the binary has higher $\vesc$ and therefore the velocities of the collisions are higher. Equations~(\ref{eq:tsh}) and (\ref{eq:lsh}) suggest that $L \propto M$, roughly similarly to what is observed in Figure~\ref{fig:sensitivity_m}. 

A striking feature is the sharp drop in $L$ at $M \approx 3\,\msun$. We have carefully verified that this feature is not a numerical artifact\footnote{The precise value of $M$ where the drop occurs is found to depend slightly on the simulation resolution and the exact procedure used to calculate spatial derivatives in our SPH equations.}, but instead represents a transition between outflows which efficiently form dust ($M \lesssim 3\,\msun$), and those in which radiative diffusion dominates ($M \gtrsim 3\,\msun$), keeping the outflow isothermal and preventing dust formation. Note that the boundary of this transition depends sensitively on the semi-major axis, and other binary properties. We investigate the thermodynamic structure of these outflows and the role of dust formation in Section~\ref{sec:dust}.

\subsection{Semi-Analytic model}

The complex range of physics (e.g., EOS, opacities) at work and their interplay with radiative diffusion and cooling makes it impossible to construct a fully analytic model that would reproduce our simulation results to better than a few orders of magnitude.  Nevertheless, to highlight the pertinent physics and to provide order-of-magnitude estimates relevant for comparing with observations, we construct a {\it semi-analytic model}, which predicts the time evolution of $L$ and $\teff$ for a given binary by making use of only the EOS and opacity tables.

\begin{figure}
\centering
\includegraphics[width=\columnwidth]{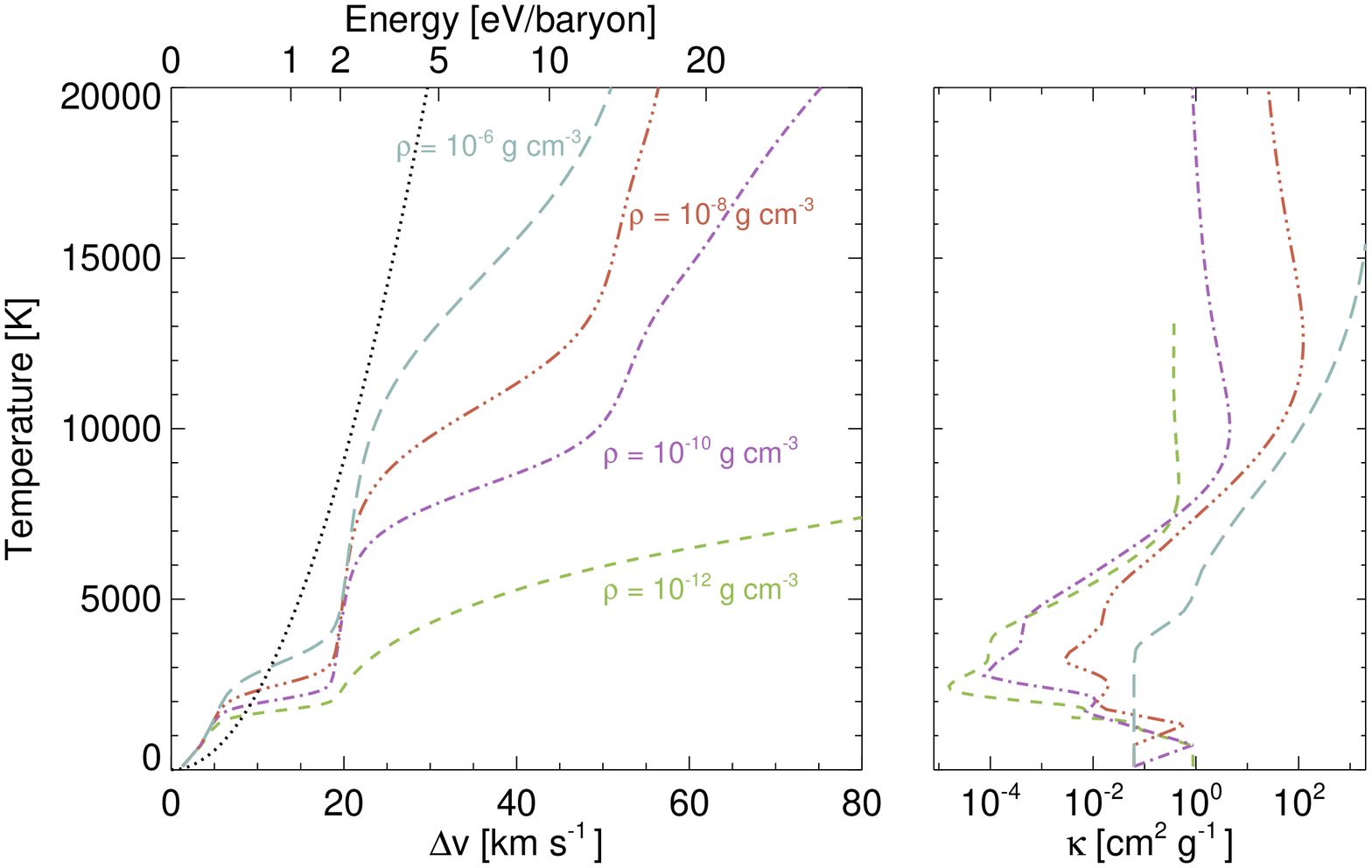}
\caption{{\em Left}: Temperature of the post-shock gas as a function of the shock velocity $\Delta v \approx \sqrt{2\Delta u}$, shown for several (fixed) densities.  The ideal gas approximation from Equation~(\ref{eq:tsh}) are shown as dotted lines. {\em Right}: Opacity corresponding to the temperatures and densities from the left panel.}
\label{fig:heating}
\end{figure}

For two streams to collide at a radius $\rc = \eta a \sim 2\eta R_*$ the velocity dispersion between them must be $\Delta v/v_{\infty} \sim 2\eta^{-1}$, where $v_{\infty} \sim 0.25 \vesc$ is the typical asymptotic velocity (Fig.~\ref{fig:leapfrog}) giving $v_\infty \approx 50\,{\rm km}\,{\rm s}^{-1}\,(m_*/r_{5*})^{1/2}$, $R_* \sim a/2$ is the radius of the primary, and $\eta \sim 8$ (Fig.~\ref{fig:structure}), as determined from the simulations\footnote{The parameter $\eta$ is related to $\varepsilon$, because we would expect $\eta \rightarrow \infty$ for $\varepsilon\rightarrow 0$, when no collisions between the streams occur. However, $\varepsilon$ describes the thermal content of the gas at the injection point, which is then subsequently modified by irradiation, adiabatic and radiative cooling, and especially the torqueing by the gravitational field of the binary. As a result, the relation between $\varepsilon$ and $\eta$ is only indirect and we keep them separate in the semi-analytic model.}.  Here the primary mass and radius are normalized to Solar values, $r_{5*} = R_*/5R_\odot$ and $m_* = M/\msun$. The gas gains energy $\Delta u \sim \Delta v^2/2$ during the collision, which raises its temperature.  Neglecting H$_2$ destruction or H ionization, the nominal shock temperature for an ideal gas would be \citep[p.~360]{lamers99}
\begin{eqnarray}
T_{\rm sh} \approx \frac{3}{16}\frac{m_{\rm p}}{k_{\rm B}} \Delta v^2 \sim 3500\,{\rm K} \left(\frac{\eta}{8}\right)^{-2} \frac{m_*}{r_{5*}} \sim  0.3\,{\rm eV} \left(\frac{\eta}{8}\right)^{-2} \frac{m_*}{r_{5*}},
\label{eq:tsh}
\end{eqnarray}
However, for many binary configurations the latent heat associated with dissociation or ionization of hydrogen is crucial for determining $T_{\rm sh}$.  In the left panel of Figure~\ref{fig:heating}, we show $T_{\rm sh}$ for a range of $\Delta v$ using our EOS and for several typical densities. $T_{\rm sh}$ is nearly constant at $\sim 2000$ to $4500$\,K for $5 \lesssim \Delta v \lesssim 20$\,\kms due to the latent heats of H$_2$ formation and destruction. Similarly, the behavior of $T_{\rm sh}$ for $20 \lesssim \Delta v \lesssim 50$\,\kms\ is affected by hydrogen ionization for $\rho \gtrsim 10^{-10}$\,g\,cm$^{-3}$.  For lower densities, radiation dominates the EOS, resulting in a weaker dependence of $T_{\rm sh}$ on $\Delta v$, as illustrated by the $\rho = 10^{-12}$\,g\,cm$^{-3}$ case shown in Figure~\ref{fig:heating}.  As a result of these complications, the correspondence between $T_{\rm sh}$ and $\Delta v$ matches Equation~(\ref{eq:tsh}) only roughly (dotted line in Fig.~\ref{fig:heating}). In the semi-analytical model, we calculate $T_{\rm sh}$ directly from $\Delta v$ using our EOS.

The vertical surface density of the outflow at radius $r$ is approximately given by
\beq
\Sigma_z \sim \frac{\mdot}{2\pi r v_\infty},
\label{eq:sigmaz_est}
\eeq 
where $v_\infty \approx 0.25\vesc$ (Figs.~\ref{fig:leapfrog} and \ref{fig:velocity}). The vertical optical depth $\tau_z \sim \Sigma_z \kappa$ through the outflow at the collision radius $\rc$ is
\beq
\tauzrc \sim \frac{\mdot\kappa}{2\pi \rc v_\infty} \approx  3 \left(\frac{\kappa}{10^{-2}\,{\rm cm}^2\,{\rm g}^{-1}}\right)(m_* r_{5*})^{-1/2} \mdot_{-3},
\label{eq:tauz_est}
\eeq
where $\mdot = \mdot_{-3} 10^{-3}$\,\myr, and the opacity is evaluated at $(T_{\rm sh},\rho_{\rm c})$, where $\rho_{\rm c}$ is the density at $\rc$,
\beq
\rho_{\rm c} \sim \frac{\Sigma_z}{2a}.
\label{eq:rhoc}
\eeq
Here we have assumed that the vertical thickness of the outflow at $r_{\rm c}$ approximately equals $a$ for $\varepsilon \sim 0.05$ (Sec.~\ref{sec:rho_t}, Fig.~\ref{fig:rz}, and Eq.~[\ref{eq:zanal}]).  Although the outflow spreads in the vertical direction, using $\rho_{\rm c}$ from Equation (\ref{eq:rhoc}) work well in the semi-analytic model.  Note that in our actual simulations we calculate $\Sigma_z$ and $\tau_z$ more accurately than in Equations~(\ref{eq:sigmaz_est}--\ref{eq:tauz_est}), as described in Appendix~\ref{app:cool}.

\subsubsection{Optically-thin outflow}

If the outflow is optically-thin in the vertical direction, $\tauzrc \lesssim 1$, then its luminosity directly matches that dissipated by the shock,
\beq
\lthin \sim \frac{1}{2}\mdot \Delta v^2 \sim 5\times 10^{34}\,\ergs  \mdot_{-3}\left(\frac{\eta}{8}\right)^{-2}m_*r_{5*}^{-1}.
\label{eq:lsh}
\eeq
When H$_2$ dissociation or H ionization can be neglected, the effective temperature of the emission is given by
\beq
\teff \sim \left(\frac{\lthin}{\sigma \pi r_{\rm c}^2}\right)^{1/4} \sim 2500\,{\rm K}\ \mdot_{-3}^{1/4} m_*^{1/4}r_{5*}^{-3/4}\left(\frac{\eta}{8}\right)^{-1},
\label{eq:Teffanalytic}
\eeq
where $\sigma$ is the Stefan-Boltzmann constant.  As shown in Figure~\ref{fig:structure} (bottom right panel), the optically-thin cooling is concentrated in a narrow half-ring of radius $\sim r_{\rm c}$ and not the full disk assumed in (eq.~\ref{eq:Teffanalytic}); $\teff$ nevertheless depends only weakly on such geometric factors.

\subsubsection{Optically-thick outflow}

\begin{figure}
\centering
\includegraphics[width=\columnwidth]{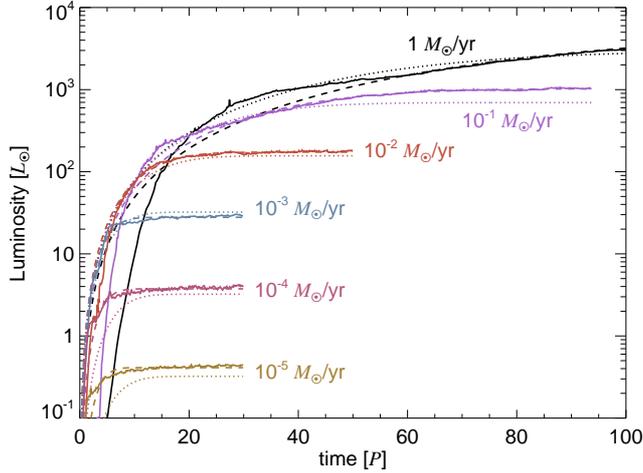}
\caption{Time evolution of the luminosity (solid lines) for simulations of a binary from Figure~\ref{fig:structure}, shown for different values of the (temporally fixed) mass loss rate $\mdot$ as marked.  Shown for comparison is the best-fit model using Equation~(\ref{eq:fit}) (dashed lines), and the predictions of the semi-analytic model (dotted lines, Eqs.~[\ref{eq:fit}--\ref{eq:tscale}]). }
\label{fig:analytic_lc}
\end{figure}

Predicting the luminosity of an optically-thick outflow with $\tauzrc \gtrsim 1$ is not as straightforward as in the optically-thin case. Because an optically-thick outflow cannot immediately radiate energy dissipated by the shock, part of the shock luminosity is lost to $P\intd V$ work as the outflow expands.  The details of this process depend on the opacities and EOS, which are not simple functions of density and temperature (Figs.~\ref{fig:eos} and \ref{fig:kappa}). We first attemped to predict the asymptotic luminosity based on simple assumptions (polytropic EOS, constant $\kappa$), but found several orders-of-magnitude discrepancies between the analytic and numerical results. We therefore take a more empirical approach based on a set of our simulation results.

For fixed binary parameters and mass outflow rate $\mdot$, the time evolution of luminosity from our simulations is well described by the functional form
\beq 
L(t) = \lthickmax\left[1 - \exp\left(-\frac{t^2}{\tscale^2} \right) \right],
\label{eq:fit}
\eeq
where $L_{\rm thick,max}$ is the asymptotic luminosity achieved at late times, and $\tscale$ is the timescale to approach the steady state. For $t \ll \tscale$, Equation~(\ref{eq:fit}) gives the expected results that the luminosity increases in proportion to the surface area of the outflow, $L \propto t^2$.  Equation~(\ref{eq:fit}) applies also to the optically-thin case, assuming we adopt a small value of $\tscale$. Figure~\ref{fig:analytic_lc} compares our simulation results for $L(t)$ to equation (\ref{eq:fit}) for several values of $\mdot$, spanning the optically-thin to optically-thick regimes.  We generally find good agreement between the simulation results and model fit.  

\begin{figure}
\centering
\includegraphics[width=\columnwidth]{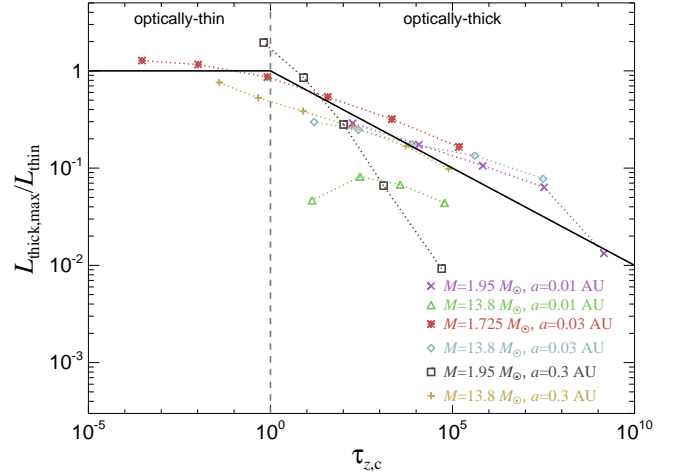}
\caption{Ratio of the steady-state luminosity to the optically-thin luminosity (Eq.~[\ref{eq:lsh}]) as a function of optical depth at the collision radius $\tauzrc$. Different symbols mark series of models with constant $M$, $q$, and $a$, but varying $\mdot$. All calculations were perfomed without the irradiation by the central binary to not bias the results due to reprocessing of the central binary radiation.}
\label{fig:analytic_model}
\end{figure}

\begin{figure}
\centering
\includegraphics[width=\columnwidth]{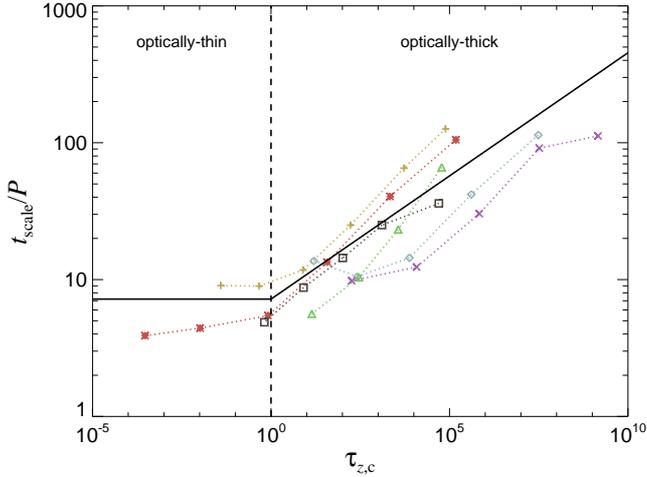}
\caption{Same as Figure~\ref{fig:analytic_model} but for the timescale to approach the steady state $\tscale$.}
\label{fig:analytic_model_scale}
\end{figure}

Figures~\ref{fig:analytic_model} and \ref{fig:analytic_model_scale} show how the parameters in Equation~(\ref{eq:fit}) depend on the key properties of the outflow. Figure~\ref{fig:analytic_model} shows the ratio of the asymptotic luminosity $L_{\rm thick,max}$ to the hypothetical optically-thin luminosity $\lthin$ from Equation~(\ref{eq:lsh}) as a function of the vertical optical depth $\tau_z(\rc)$.  For $\tauzrc \lesssim 1$ we obtain $\lthickmax/\lthin \sim 1$  as expected, but for $\tauzrc \gtrsim 1$ the asymptotic luminosity is suppressed with increasing $\tauzrc$, approximately as a power-law. This behavior is well fit by the functional form,
\beq
\lthickmax \sim \lthin \left[\max (1,\tauzrc) \right]^{-0.20}
%\frac{\lthickmax}{\lthin} \sim \tauzrc^{-0.20}.
\label{eq:lsh_thick}
\eeq
Although the value of the power-law index ($-0.2$) is not obvious from any order-of-magnitude estimates, this is not surprising given the complex interplay in our simulations between radiative diffusion and cooling, and the properties of the EOS and opacities. The scatter about the best-fit power-law is smaller than order of magnitude, which indicates that our semi-analytic model captures the pertinent processes with good accuracy using Equation~(\ref{eq:lsh_thick}). Part of the scatter comes from the difference between the diffusion-dominated and cooling-dominated outflows described in Section~\ref{sec:dust}. We emphasize that our fit to $L_{\rm thick,max}(\tauzrc)$ factors in the size and mass of the binary, as well as the mass-loss rate, each of which varies by orders of magnitude across the range of simulated stellar systems.

Figure~\ref{fig:analytic_model_scale} shows how the characteristic timescale to reach steady state $\tscale$ depends on $\tauzrc$. As with $L_{\rm thick,max}$, the $\tscale$ is approximately constant for $\tauzrc \lesssim 1$, while increasing as a power law for $\tauzrc \gtrsim 1$.  This behavior is well fit by
\beq
\tscale \sim \frac{2\rc}{v_\infty} \left[\max(1,\tauzrc)\right]^{0.18},
\label{eq:tscale}
\eeq
across all $\tauzrc$.  Although the scatter along Equation~(\ref{eq:tscale}) is again less than an order-of-magnitude, this can introduce potentially greater inaccuracies when trying to predict $L(t)$ for $t \lesssim \tscale$ from Equation~(\ref{eq:fit}).

In Figure~\ref{fig:analytic_lc}, we show the predicted light curve calculated from our semi-analytic model ( equations~[\ref{eq:fit}]--[\ref{eq:tscale}]) in comparison to the actual simulation results.  The agreement is relatively good.  For other binary parameters, the agreement can be noticeably worse, for example due to time-dependent dust formation effects (Sec.~\ref{sec:dust}).  Nevertheless, the semi-analytic model typically matches the simulations comparable to or better than order-of-magnitude for $t \gtrsim \tscale$. 

\begin{figure}
\centering
\includegraphics[width=\columnwidth]{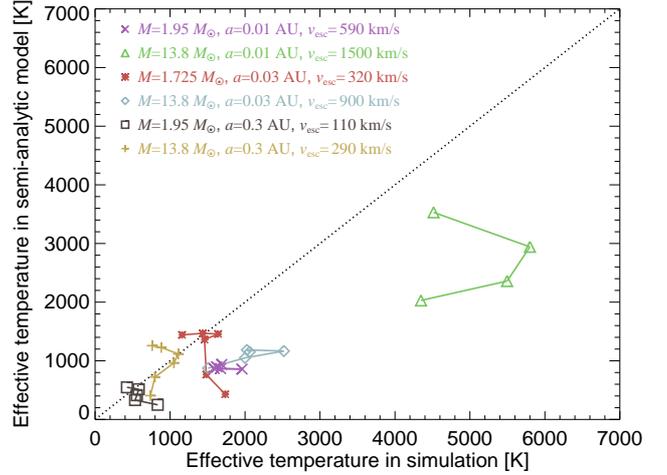}
\caption{Comparison of asymptotic effective temperatures measured in the simulations and predicted by the semi-analytic model (Eq.~[\ref{eq:teff_thick}]). The symbols are the same as in Figures~\ref{fig:analytic_model} and \ref{fig:analytic_model_scale}.}
\label{fig:teff_comp}
\end{figure}

Finally, we estimate the asymptotic effective temperature at late times $t \gg \tscale$ in the optically-thick case as
\beq
\teff^4 \sim \frac{L_{\rm thick,max}}{4\pi (v_\infty \tscale)^2 \stefb}.
\label{eq:teff_thick}
\eeq
In Figure~\ref{fig:teff_comp}, we compare the effective temperatures from the simulations to those predicted by Equation~(\ref{eq:teff_thick}).  The semi-analytic model underpredicts the temperatures, especially for $\teff \gtrsim 3000$\,K, an important fact to keep in mind when comparing our semi-analytic model to observations (Sec.~\ref{sec:summary}).

\subsubsection{Extensions of the model}
\label{sec:extension}
\begin{figure}
\centering
\includegraphics[width=\columnwidth]{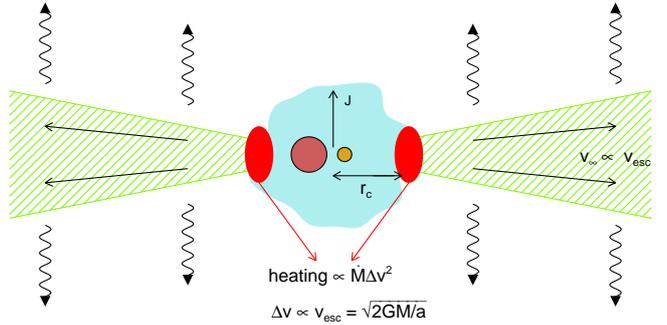}
\caption{ Schematic configuration of mass-losing binary stars, where our numerical and semi-analytic results describe the resulting electromagnetic emission. Binary star with angular momentum vector $\mathbf{J}$ and escape velocity $\vesc=\sqrt{2GM/a}$ produces an equatorially-concentrated outflow with an asymptotic velocity $v_\infty \propto \vesc$, which is heated at a distance $\rc$ from the binary with a rate $\mdot \Delta v^2$, where $\Delta v \propto \vesc$. The radiation escapes the outflow primarily in the vertical direction.}
\label{fig:diagram}
\end{figure}

The semi-analytic model presented in this Section can be easily extended to a more general class of models relevant for the dynamic shrinking of the binary orbit inside of a common envelope. The semi-analytic model describes a disk-like primarily radial outflow with asymptotic velocity $v_\infty$, which is heated at the inner edge ($r \sim \rc$) at a rate $\lthin$ (Eq.~[\ref{eq:lsh}]), as shown in Figure~\ref{fig:diagram}. The physics of \ltwo\ mass loss determines the structure of the outflow inside of $\rc$, which sets the relative ratio of the outflow velocity to the binary escape velocity, $v_\infty/\vesc \approx 0.25$ (Fig.~\ref{fig:leapfrog}), the relation between $\vesc$ and the heat deposition rate, $\Delta v/\vesc \approx 0.5/\eta$, and the radius $\rc$ where the heat is deposited. It is reasonable that binary inspiral inside a common envelope will also lead to an equatorially-concentrated outflow, with a velocity and heating rate near its inner edge which are proportional to the instantenous escape speed of the binary \citep[e.g.][]{taam89,livio90}. Naturally, the precise numerical values of $v_\infty/\vesc$, $\Delta v/\vesc$, $\rc/a$, and even the functional form of $\lthin(\vesc)$ will likely be different, and potentially varying in time, but the observable features will be described by a similar semi-analytic theory.  Specifically, the shock heating of the ejected material might happen much closer at $\rc \approx a$. This should not change the outcome qualitatively, because a large portion of the energy will be lost to adiabatic expansion before the material can efficiently radiate, as shown in our work.

\section{Dust formation}
\label{sec:dust}

A necessary condition for dust formation is that the outflow temperature decrease below the dust condensation temperature, $\tdust$. Although this is the only condition to form dust within the \fa\ opacity table (where we estimate $\tdust \approx 1200$\,K from the opacity increase; Fig.~\ref{fig:kappa}), this condition is not sufficient. Following \citet{kochanek11b}, the maximum size of dust grains, $a_{\rm max}$, which can form at a given radius $r_{\rm dust}$ is
\beq
a_{\rm max} = \frac{v_{\rm dust}X_{\rm g}\mdot}{16\pi (z_{\rm max}/r_{\rm dust})\rho_{\rm bulk}v_\infty^2r_{\rm dust}},
\eeq
where $v_{\rm dust} \approx 1$\,\kms\ is the collision velocity between coagulating particles for gas of temperature $\approx \tdust \sim 10^{3}$ K, $X_{\rm g}=0.005$ is the mass fraction of condensible material, $\rho_{\rm bulk} \sim 3$\,g\,cm$^{-3}$ is the bulk density of the grain, $z_{\rm max}$ is the vertical thickness of the outflow, and $v_\infty \approx 0.25\vesc$. The condition to form dust is $4a_{\rm max}^3\rho_{\rm bulk}/m_0 > N_{\rm dust}$, where $m_0 = 12\mpr$ and $N_{\rm dust} \sim 7$ is some critical number of molecules to start the coagulation process. This corresponds to a mass-loss rate of
\beq
\mdot \gtrsim 10^{-9}\,\myr\, m_* \left(\frac{r_{\rm dust}}{\rc}\right)\left(\frac{\eta}{8}\right) \left(\frac{z_{\rm max}/r_{\rm dust}}{0.1}\right).
\label{eq:mdot_dust}
\eeq
Equation (\ref{eq:mdot_dust}) shows that cool \ltwo\ outflows can in principle satisfy the condition for dust grain growth, even at very low $\mdot$.  This conclusion is relatively insensitive to the semi-major axis of the binary.  We are thus justified to use the \fa\ opacity tables, which include dust grains.

\begin{figure}
\centering
\includegraphics[width=\columnwidth]{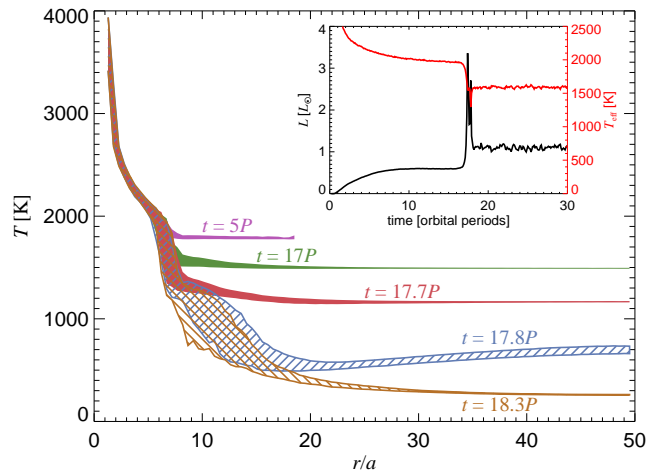}
\caption{Evolution of a radial dependence of temperature for the binary from Figure~\ref{fig:structure} with $\mdot=10^{-5}$\,\myr for five epochs after the start of the simulation. The shaded and hatched regions denote the $5\%$ and $95\%$ of the temperature distributions as a function of radius. The inset plot shows time evolution of the luminosity and effective temperature produced by the outflow. For clarity, this calculation does not include the irradiation by the central binary. }
\label{fig:transition}
\end{figure}

We have discovered another effect that can suppress dust formation in our simulations. In Figure~\ref{fig:transition}, we show the temperature as a function of radius and time for the SG case with $\mdot=10^{-5}$\,\myr. The heat generated by the spiral shocks is initially efficiently redistributed radially by radiative diffusion, because $\dot{u}_{\rm diff}$ dominates the energy equation.  This results in the outflow being nearly isothermal for $r > \rc \approx 8a$ ($t=5P$ in Fig.~\ref{fig:transition}). As the outflow evolves, however, a larger fraction of the power is lost to adiabatic expansion, causing the temperature of the isothermal outflow to gradually decrease with time ( until $t=17P$ in Fig.~\ref{fig:transition}). When the outflows approaches the dust formation temperature, $\kappa$ dramatically increases and radiative cooling becomes the dominant term in the energy equation. The outflow temperature suddenly decreases as a result, essentially on the diffusion timescale ($t=17.8P$), and a new equilibrium corresponding to a dusty outflow is quickly established ($t=18.3P$). 

The inset plot illustrates this behavior by showing the time evolution of the radiative luminosity and effective temperature. The luminosity initially rises as the shocked outflow is established, but then asymptotes to a constant value as the outflow becomes cooler as described above.  When the $T \approx \tdust$ and dust first forms, the luminosity spikes, releasing the accumulated thermal energy.  There is an associated decrease in the effective temperature, because cooler areas of the outflow now contribute to the luminosity as a result of the higher value of $\kappa$.

\begin{figure}
 \centering
\includegraphics[height=5.5cm]{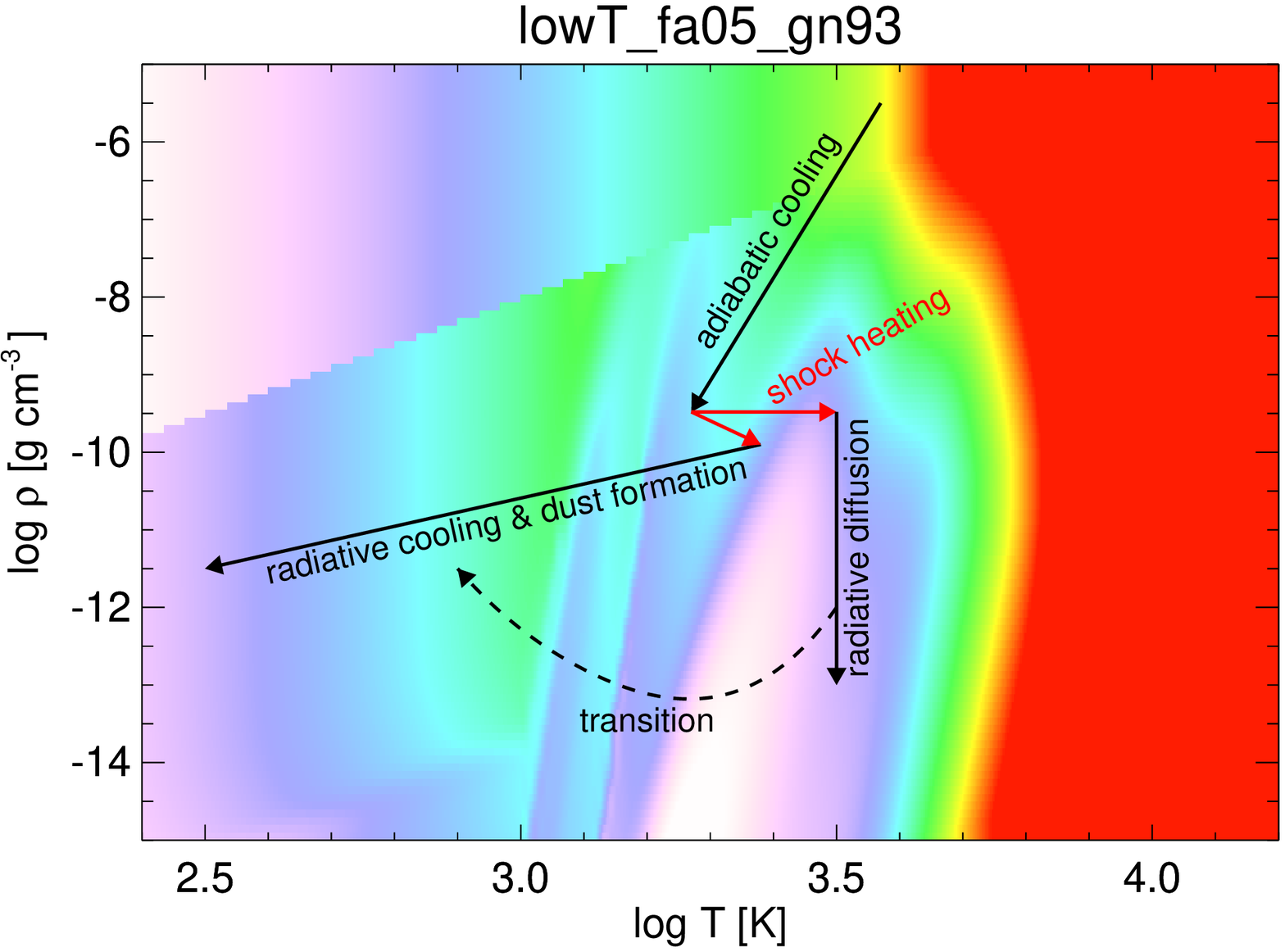}\includegraphics[height=5.5cm]{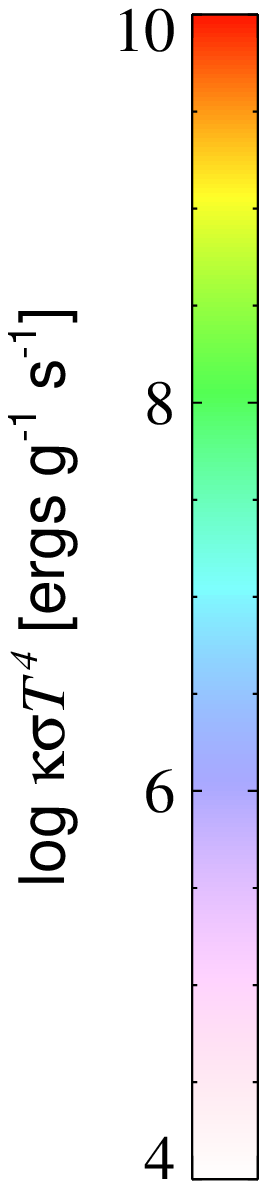}
\caption{Schematic illustration of the origin of the two types of outflows (Fig.~\ref{fig:sensitivity_m} and \ref{fig:transition}) and their transition in temperature-density space, with the optically-thin cooling rate $\kappa \stefb T^4$ as a shaded background.  The arrows represent particle densities and temperatures from snapshots of our simulations. Initially, the gas cools quasi-adiabatically along lines of decreasing temperature and density, until being heated when the encounter the shock.  If the shock heating is weak, then the gas remains at low temperatures and high opacities; dust forms and the resulting increased opacity and radiative cooling rate drive the outflow to even lower temperatures. Alternatively, if shock heating is stronger and places the gas into the $\rho-T$ region of low radiative cooling efficiency, then radiative diffusion dominates and drives the outflow to a radially constant temperature.  Due to adiabatic losses, the temperature of the isothermal outflow also slowly decreases as the outer radius $r_{\rm out}$ grows in time.  Once the dust formation temperature is reached, the entire outflow abruptly transitions to the former dust-dominated state.  See the text for more details.}
\label{fig:bifurcation}
\end{figure}

Figure~\ref{fig:bifurcation} further illustrates this scenario by showing the temperature and density evolution of the outflow on a background shaded by the optically-thin cooling rate.  After leaving the \ltwo, the gas expands freely along trajectories of decreasing temperature and density.  Upon encountering the previous wrapping of the spiral, it is heated to the temperature $T_{\rm sh}$ (Eq.~[\ref{eq:tsh}]).  Depending on the binary parameters ($M$, $a$) and the outflow density, the density and temperature can land in the opacity valley ($3.2 \lesssim \log T \lesssim 3.6$, $\rho \lesssim 10^{-9}$\,g\,cm$^3$), where the optically-thin cooling rates are several orders of magnitude smaller than in the surrounding areas. 

The energy equation (Eq.~[\ref{eq:u}]) is dominated by radiative diffusion, which drives the outflow to constant temperature at $r > \rc$.  As the gas loses energy to adiabatic expansion (and to a lesser degree also to radiative cooling), this radially-constant temperature decreases with time (Figure~\ref{fig:transition}). The rate of adiabatic energy loss for $r > \rc$, where $\rho \propto r^{-2}$, is
\beq
\dot{u}_{\rm gas} \sim -2\frac{P}{\rho} \frac{v_\infty}{r},
\eeq
where $v_{\infty} \propto \vesc$ is the asymptotic velocity (assumed to be constant for $r>\rc$). For an ideal gas with $P = \kb \rho T/\mpr$, the total adiabatic energy loss between $\rc$ and the outer radius of the outflow, $r_{\rm out}$, is thus given by
\beq
\int_{\rc}^{r_{\rm out}}\dot{u}_{\rm gas} \frac{\mdot}{v_\infty}\intd r = -2\frac{\kb T}{\mpr}{\mdot}\ln\left( \frac{r_{\rm out}}{\rc}\right).
\eeq
The total available power compensating for adiabatic losses is temporally constant and equal to $\lthin$ (Eq.~\ref{eq:lsh}]), providing a relationship between the outflow temperature $T$, $r_{\rm out}$, and the velocity spread of the spiral $\Delta v^2$.  Once $T \rightarrow \tdust$, radiative cooling suddenly becomes much more efficient, and the outflow radiates its thermal energy on the radiative diffusion timescale. This happens once the outflow raches a critical size
\beq
\ln\left(\frac{r_{\rm out}}{\rc}\right) \sim \frac{1}{4}\frac{\mpr \Delta v^2}{\kb \tdust} \sim \frac{4}{3} \frac{T_{\rm sh}}{\tdust}.
\label{eq:rout}
\eeq
For higher values of $T_{\rm sh}$, an exponentially larger outer radius is required to reach the dust formation temperature $\tdust$.  For the example shown in Figure~\ref{fig:transition}, $T_{\rm sh} \approx 1900$\,K ($t=5P$ in Fig.~\ref{fig:transition}) and $\tdust \approx 1200$\,K, which gives $r_{\rm out} \sim 8\rc$. The first gas element requires a time $t \sim 30P$ to reach this $r_{\rm out}$, a result in broad agreement with Figure~\ref{fig:transition}. Note that because $r_{\rm out}$ is exponential in $T_{\rm sh}$, small changes in the binary parameters results in dramatic changes in the time of the dust transition and the amount of energy released.

The total amount of energy released when dust forms is given by
\begin{eqnarray}
E &\sim& \frac{3}{2}\frac{\kb\tdust}{\mpr} \frac{\mdot}{v_\infty} \eta a \exp\left(\frac{4}{3}\frac{T_{\rm sh}}{\tdust}\right) \sim \nonumber\\
&\sim&  10^{41}\,{\rm erg} \left(\frac{\tdust}{1200\,{\rm K}}\right) \mdot_{-3}\frac{r_{5*}^{3/2}}{m_*}\left(\frac{\eta}{8}\right)\left(\frac{r_{\rm out}/\rc}{10}\right).
\end{eqnarray}
The radiative diffusion timescale across the outflow is
\beq
t_{\rm diff} \sim \frac{r_{\rm out}^2 \kappa\rho}{c} \sim 16\,{\rm s}\  \left(\frac{r_{\rm out}}{10\rc}\right)r_{5*}^2 \left(\frac{\kappa}{10^{-4}\,{\rm cm}^2\,{\rm g}^{-1}}\right)\rho_{-10},
\eeq
where $\rho = \rho_{-10} 10^{-10}$\,g\,cm$^{-3}$ and we assumed typical opacity for the opacity minimum. As the radiative diffusion timescale is short in all cases of interest, the duration of the luminosity spike is limited by the time to radiate the thermal energy,
\beq
t_{\rm cool} = \frac{u}{\dot{u}_{\rm cool}} \sim 10^4\,{\rm s}\ \left(\frac{\tdust}{1200\,{\rm K}}\right)^{-3}\left(\frac{\kappa}{10^{-1}\,{\rm cm}^2\,{\rm g}^{-1}}\right)^{-1},
\eeq
where we used typical opacity for dust grains. Alternatively, the finite time required to form dust grains could be the rate limiting step in the rapid transition (eq.~[\ref{eq:mdot_dust}]). The double-peaked structure in the time evolution of $L$ in Figure~\ref{fig:transition} is caused by changes in $\kappa$ as $T$ decreases (Fig.~\ref{fig:bifurcation}).

Note also that the transition between radiation diffusion dominated outflow and dust formation will appear differently from different viewing angles.  As dust formation typically starts at $r_{\rm out}$ and then propagates inward, the binary will become optically-thick in the orbital plane almost immediately, causing the luminosity to drop significantly.  However, the outflow can remain optically-thin in the vertical direction, where all the radiation escapes, and so after dust forms the binary will continue to be bright when viewed approximately pole-on.

There are reasons to believe that the sudden formation of dust observed in our simulations might be an artifact of the simplifications made in this work. First, radiative diffusion should drive the system to a constant radiative flux (a gradient of radiation energy), rather than radially constant temperature. However, the \ltwo\ outflow is not thermally supported and has a finite outer boundary, because the particles are being injected into an empty space. When radiative cooling is inefficient, radiative flux cannot leave the system at the outer boundary, so the only way to have radially constant flux is with zero flux, i.e. a radially constant temperature. Also, in physically realizable situations, the \ltwo\ outflow does not begin instantaneously, but instead grows on many dynamical timescales of the binary, as was potentially observed in V1309~Sco \citep{pejcha14}. In such cases, enough previous matter might be already present to enable radiation to quickly diffuse out, which will then not present an obstacle to dust formation. 

Second, we do not differentiate between the Planck and Rosseland opacity means in the expressions for radiation cooling. In the region of low opacity, the Planck means might be noticeably higher than the Rosseland means \citep[Fig.~27]{tomida13}. As a result, optically-thin radiative cooling will not be as strongly suppressed and might dominate the energy equation, again suppresing the diffusion and leading to a rapid dust formation. Finally, the properties of the dust such as the opacity depend on the metallicity of material, which we assumed to be constant and equal to approximately solar. Furthermore, the composition of dust in stellar mergers might be very different from the normal dust in terms of isotopic ratios \citep{kaminski15}. To summarize, our results indicate only the possibility rather than a certainty of sudden dust formation. This issue should be illuminated with simulations employing more sophisticated radiation transport, including more appropriate opacity tables.

\section{Summary of Observable Properties and Connection to Red Transients}
\label{sec:summary}

\begin{figure*}
\centering
\includegraphics[width=\textwidth]{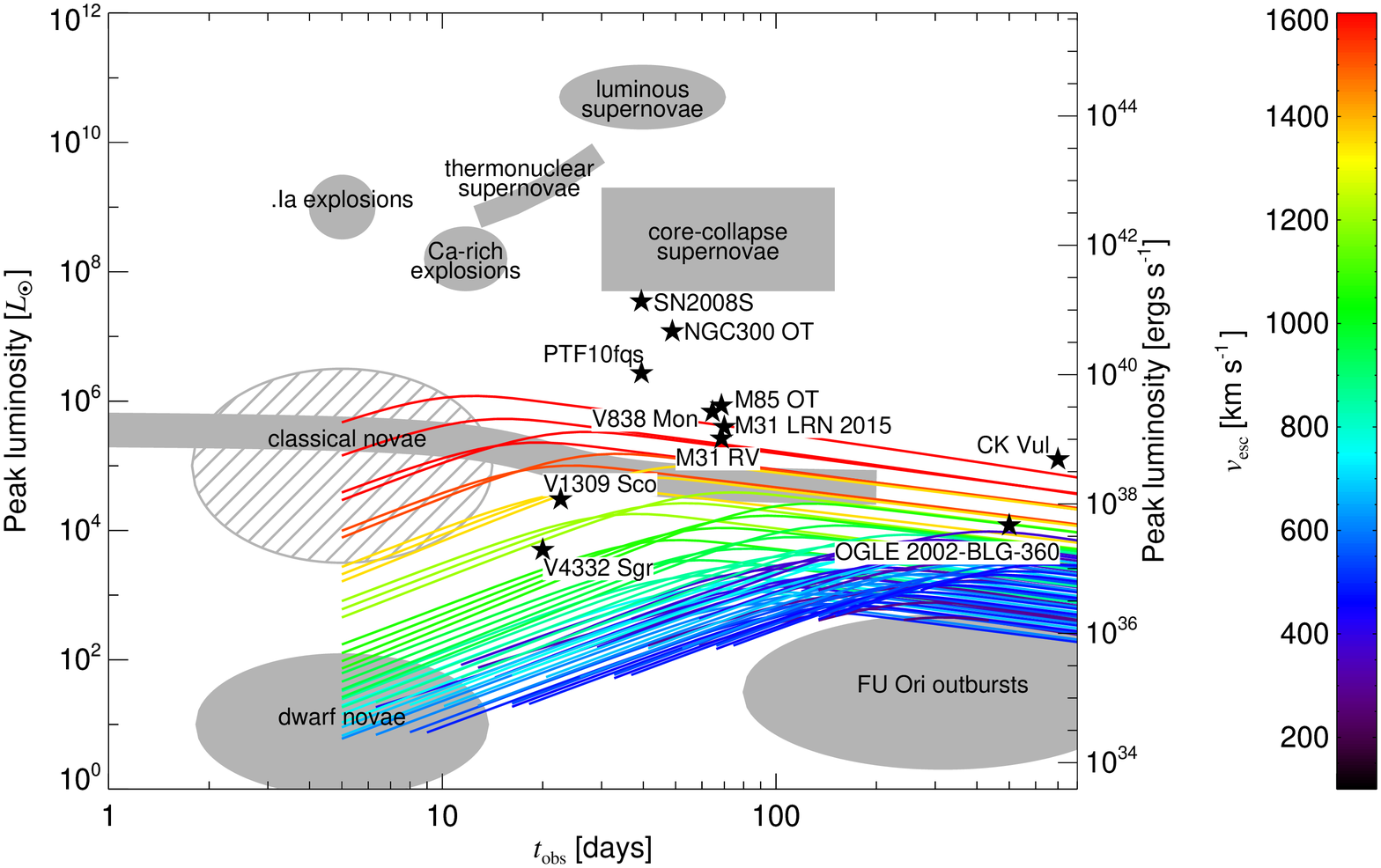}
\caption{Peak luminosities and durations of astronomical transients after \citet{kasliwal11,kasliwal12}. In addition to objects and classes in \citet{kasliwal11,kasliwal12}, we show dwarf novae \citep{warner03}, FU~Ori outbursts \citep{hartmann89,rodriguez90}, and red transients V4332~Sgr \citep{tylendaetal05a}, V1309~Sco \citep{tylenda11}, M31 LRN 2015 \citep{williams15}, OGLE~2002-BLG-360 \citep{tylenda13}, and CK Vul \citep{kato03}. Colored lines are predictions for mass-losing binaries from our semi-analytic model, where each line corresponds to a single set of binary parameters and a range of timescales (left as a free parameter).  The color marks the escape velocity of the binary, as denoted in the colorbar on the right.}
\label{fig:summary}
\end{figure*}

We have shown that the luminosity generated by the partial dissipation of the kinetic energy of the outflow from the \ltwo\ point is directly proportional to the mass-loss rate and the escape velocity of the binary $\vesc = \sqrt{2GM/a}$. We have also argued that similar observational characteristics will be exhibited by a much wider class of models motivated by inspiral of a binary star within a common envelope: disk-like outflow with asymptotic velocity and energy deposition rate at the inner edge proportional to $\vesc$.  We now use our results to predict the observable characteristics of optical/infrared transients produced by binary mass loss.

The primary characteristics of a transient event are its peak luminosity and characteristic duration $t_{\rm obs}$ \citep{kasliwal11,kasliwal12}. If angular momentum loss through the \ltwo\ point dominates the evolution of the merging binary, the timescale of the associated transient is set by the response of the envelope of the primary star to mass loss and the associated angular momentum loss. At later phases, drag forces in the non-corotating common envelope will determine the inspiral and the energy deposition rates. These processes are challenging to predict with confidence, thus leading us to instead consider a range of timescales as a free parameter and ask what is the maximum luminosity produced by a particular binary on this timescale. We assume that the binary uniformly loses $10\%$ of its mass over the transient duration $\tobs$, resulting in a the mass-loss rate
\beq
\mdot = 0.1\frac{M}{\tobs}.
\label{eq:mdot_obs}
\eeq
We also assume that the binary parameters are approximately constant during the transient. In reality, $\mdot$ and the binary parameters must evolve during the event, which will affect the observed characteristics. Furthermore, the observed properties of the transient are expected to be viewing angle dependent, which we discuss later in more detail. Our goal here is to show that our results can broadly match some of the general characteristics of the red transients, and we defer the application our results to specific transients to forthcoming work. Thus, we estimate the transient peak luminosity as $L(\tobs)$, where $L(t)$ is evaluated using the semi-analytic model (Sec.~\ref{sec:an}). 

In Figure~\ref{fig:summary}, we show the resulting maximum luminosity as a function of the timescale for a grid of binaries binaries with $0.005 \le a \le 0.5$\,AU and $1.4 \le M \le 14\,\msun$. The lower limit of $a$ is set approximately by the size of the core of evolved massive stars. For each set of binary parameters we investigate $\tobs$ between $5$ and $800$\,days, which roughly corresponds to the observed timescales of red transients. We show only results in cases where the peak transient luminosity is at least $10$ times higher than the intrinsic luminosity of the binary $L_*$. We see that for each binary, the peak luminosity initially increases as $\tobs$ gets longer, because the luminosity is limited by the time it takes the radiation to diffuse out of the outflow, which is set by the expansion timescale of the outflow ($\tobs \lesssim \tscale$). Eventually, the peak luminosity starts decreasing with further increases in $\tobs$, because the mass-loss rate corresponding to the given timescale cannot produce sufficiently high luminosity ($\tobs \gtrsim \tscale$). The individual lines are relatively well-ordered by the escape velocity of the binary with more compact binaries (smaller $M/a$) producing higher luminosities.

Figure~\ref{fig:summary} shows that the peak luminosity reachable within our framework is $\sim 10^6\,\lsun$. The transients occupy the parameter space between dwarf novae, FU Ori outbursts, core-collapse supernovae and the bulk of classical novae.  Note that our results hold even for mechanisms for producing outflows from binaries other than \ltwo\ mass loss, unless the model parameters $\Delta v/\vesc$ and $v_\infty/\vesc$ differ by more than order of magnitude from what we assumed.  

The predictions of our semi-analytic model match the luminosities and effective temperatures of the observed class of red transients, with the exception of SN2008S, NGC300 OT, and probably also PTF10fqs. Reaching luminosities necessary to explain these events would require noticeably higher values of $\vesc$, implying a very small semi-major axis. This might not be realistic for transients involving massive stars such as SN2008S and NGC300 OT, which had progenitor masses of $6$--$10\,\msun$ \citep{prieto08,prieto09}. Note also that the timescales of optimal energy release of each binary (maxima of the lines of the semi-analytic model in Fig.~\ref{fig:summary}) decrease as $\vesc$ increases, which implies that the predicted timescales might be too short if one attempts to reproduce the luminosity of SN2008S and NGC300 OT by increasing $\vesc$. Leaving the luminosities and timescales aside, SN2008S and NGC300 OT differ in a number of other fundamental aspects from the rest of RT, as we detailed in the Introduction.

\begin{figure}
\centering
\includegraphics[width=\columnwidth]{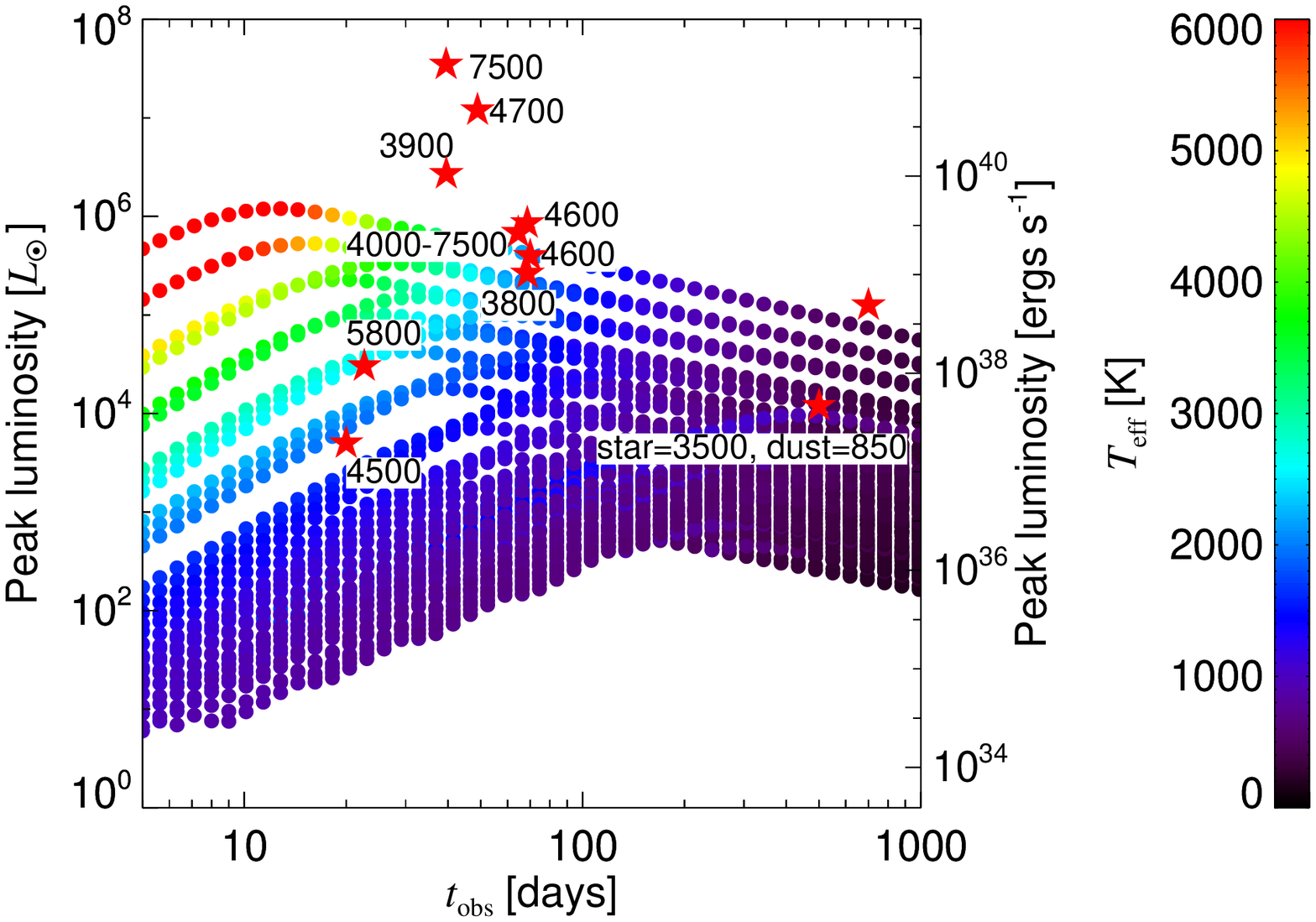}
\caption{Effective temperatures of the transients produced by the \ltwo\ mass loss as a function of the peak luminosity and the transient timescale. We also show observed red transients labeled by their peak effective temperature in Kelvin, including V1309~Sco \citep{tylenda11}, V4332~Sgr \citep{tylendaetal05a}, PTF10fqs \citep{kasliwaletal11b}, M85 OT \citep{kulkarni07}, SN2008S \citep{smith09}, NGC300 OT \citep{berger09}, V838~Mon \citep{tylenda05}, OGLE~2002-BLG-360 \citep{tylenda13}, M31 RV \citep{mould90}, and M31 LRN 2015, where we converted the F5I spectral type \citep{kurtenkov15}  to temperature using observations of \citet{vanbelle99}. The effective temperature for CK~Vul was not measured. }
\label{fig:summary_teff}
\end{figure}

In Figure~\ref{fig:summary_teff}, we show the predicted effective temperatures of our semi-analytic model in comparison with those of the known red transients. Our model generally predicts transients with low temperatures, $\teff \lesssim 3000$\,K, except the fastest and brigtest objects, which might reach up to $\sim 6000$\,K.  Note that our semi-analytic model underpredicts the effective temperature for $\teff \gtrsim 3000$\,K, which corresponds to the upper left part of Figure~\ref{fig:summary_teff}. Our $\teff$ roughly agree with most red transients, again except for SN2008S and NGC300 OT. We also predict that longer transients will have lower temperatures, potentially reaching below $1000$\,K, as was observed in OGLE~2002-BLG-360. The effective temperature of CK~Vul is unknown, but might have been higher, because the object exhibited three faster brightenings over $\sim 700$\,days. The very low effective temperatures allowed in our model are a potential distinguishing feature from pure hydrogen recombination models \citep{ivanova13b}.

We predict transient luminosities that decrease rather gradually with increasing timescale, which suggests the possibility $\sim 10^5\,\lsun$ transients with durations of several years with $\teff \lesssim 1000$\,K. As a result of their low optical luminosities, such transients may best be discovered by surveys of nearby galaxies at near-infrared wavelengths.  Indeed, such a class of transients may have recently been discovered by {\it Spitzer} \citep{kasliwal14}.

We expect the spectra of the transients to show cool photospheres appropriate for their effective temperatures with emission lines of Balmer series, and potentially rotational-vibrational transitions of H$_2$ and CO in the near-infrared. If dust is reprocessing the radiation, we expect low-temperature continuum with dust emission features. Some features of the infrared spectra might be similar to innermost regions of proto-stellar disks due to similarities in densities and temperatures \citep[e.g.][]{carmona11,beck12}.

\begin{figure*}
\centering
\includegraphics[width=0.8\textwidth]{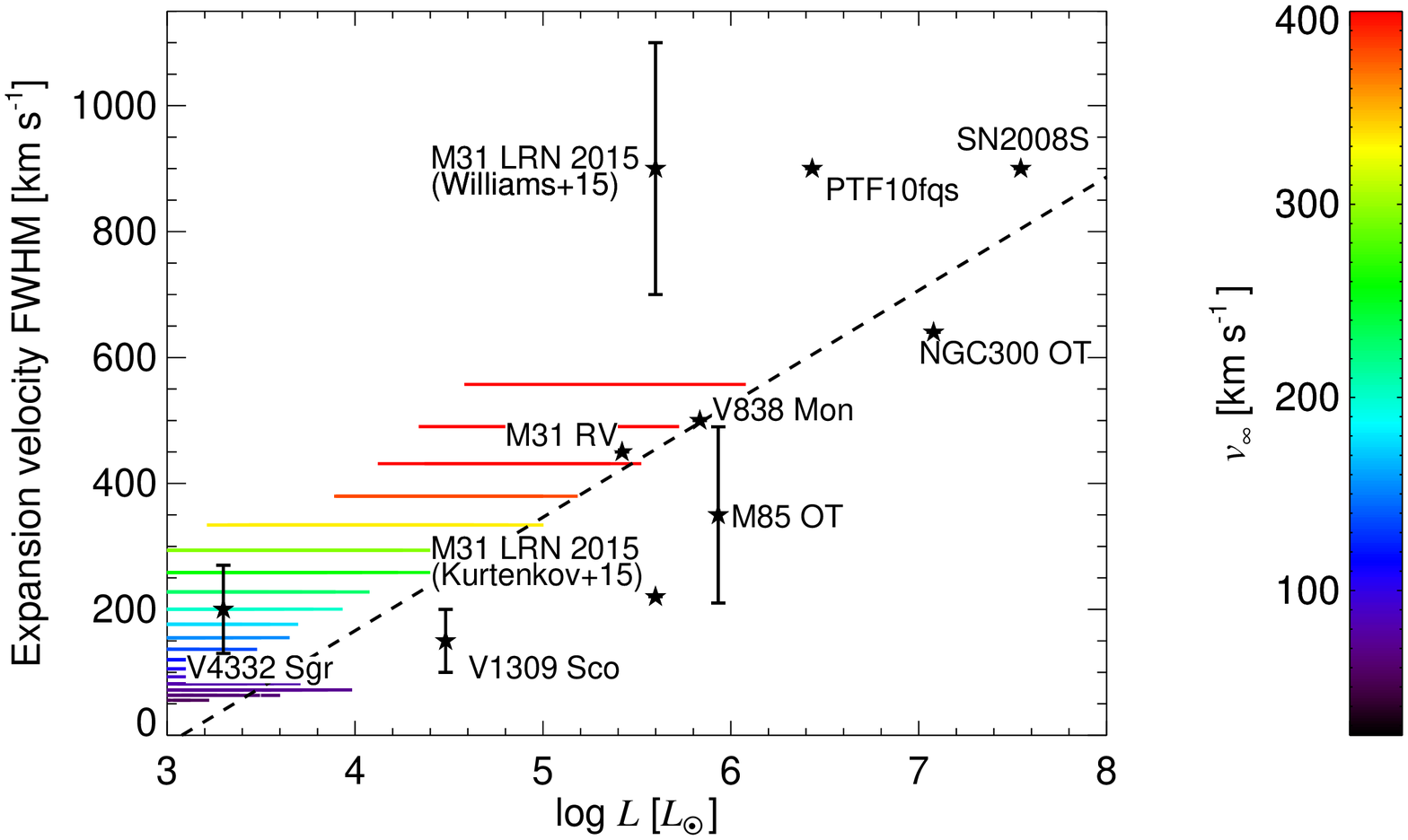}
\caption{Correlation between the expansion velocity and peak luminosity of luminous red novae. Expansion velocities were typically measured on H$\alpha$ and include uncertainties where available. We show data for V4332 Sgr \citep{martini99}, V1309 Sco \citep{mason10}, M31 RV \citep[expansion velocity obtained from Fig.~3 of][]{rich90}, V838~Mon \citep{munari02}, M85~OT \citep{kulkarni07}, PTF10fqs \citep{kasliwaletal11b}, NGC300~OT \citep{berger09}, and SN2008S \citep{botticella09}. For M31 LRN 2015 we show expansion velocity of \citet{williams15} not corrected for instrumental resolution (Williams, 2015, private communication), and of \citet{kurtenkov15}. The dashed line shows unweighted fit to the data, excluding the \citet{williams15} measurements, giving $\log (L/\lsun) = (180 \pm 50)(v_\infty/\kms) - (550 \pm 290)$.}
\label{fig:vej}
\end{figure*}

In our model the asymptotic velocity of the outflow, as well as its heating rate from inter-stream shocks, increases with the binary escape velocity. This behavior results naturally from \ltwo\ mass loss (Figs.~\ref{fig:leapfrog} and \ref{fig:summary}), but it may extend more generally to other mechanisms of binary mass loss powered by orbital decay.  This scaling also results in a correlation between the observed luminosity and the expansion velocity of the outflow. In Figure~\ref{fig:vej}, we show the asymptotic outflow velocities of our model in comparison to the observed expansion velocities of red transients, as typically derived from full-width-half-maximum of Balmer lines. The observed transients exhibit rough correlation between these two quantities, which to our knowledge was not demonstrated before. The velocities predicted by our model roughly parallel the best-fit line through the data, although the model is systematically higher. In some cases, this can be attributed to inclination effects, when the disk-like outflow is not observed exactly edge-on. In V1309~Sco, the binary was observed nearly edge-on \citep{tylenda11}, and thus part of the differences might be attributed also to uncertainties in our model, differences in the extraction of expansion velocity, and correction for the instrumental resolution of the spectrograph, which is crucial for low-resolution spectra. 

\citet{ivanova13b} proposed that the red transients are powered by the hydrogen recombination energy from an expanding shell ejected from the merging binary. Analytic estimates of the properties of transients produced by diffusion of thermal energy through a recombination front, as is commonly applied to Type II-Plateau supernovae \citep[e.g.][]{arnett80,popov93,kasen09}, predict that the ejecta velocity should be proportional to the explosion energy $E_{\rm exp}$ and ejecta mass $M_{\rm ej}$, $v_\infty \sim (E_{\rm exp}/M_{\rm ej})^{1/2}$. For a supernova, the explosion energy is set by the explosion mechanism in the inner $\sim 1000$\,km of the star and hence there is a great difference between $v_\infty$ and $\vesc$. For example, assuming that $v_\infty$ of a supernova is a fraction of $\vesc$ from the surface of a red supergiant would give completely misleading results. This illustrates the difference between an explosion and a wind-like outflow. For red transients the assumption, that $v_\infty \propto \vesc$ gives reasonable results. Both \ltwo\ mass loss model presented here, and the model proposed by \citet{ivanova13b}, yields a correlation between $v_\infty$ and peak luminosity.

\citet{kochanek14} found that the peak luminosity of the transient is a steep function of the progenitor mass, $L \propto M^{2-3}$. We would expect more massive progenitors to produce more luminous events for both our model and the hydrogen recombination model. The exact scaling with mass will depend on a combination of ejected mass, ejection timescale, the diffusion time through the ejecta, and the relation between mass and size of the binary.

We note that, due to the disk-like geometry of the outflow, which we have characterized quantitatively for \ltwo\ mass loss (Fig.~\ref{fig:rz}), the transient properties may depend on the viewing angle of the observer with respect to the orbital plane.  As $\mdot$ increases, the outflow becomes optically-thick first in the orbital plane and, if dust forms, radiation will escape primarly in the perpendicular direction.  If the orbital plane is viewed face-on, dust acts primarily to decrease $\teff$ and to prolong the timescale for radiation to  escape the outflow. If dust forms during the transient, potentially through the mechanism described in Section~\ref{sec:dust}, then observers viewing the event close to the orbital plane will observe fading and dust formation.  However, face-on observers may observe little change in $L$ and $\teff$, or even a modest brightening as rapid cooling releases the accumulated thermal energy (Fig.~\ref{fig:transition}).

Finally, it is clear that the full range of phenomenology exhibited by red transients are more complex than can be accommodated by our model, and may involve multiple mass ejection mechanisms with a geometry more complex than the equatorial outflow heated at the inner edge studied here \citep[e.g.][]{thompson09,kochanek11a,nandez14,kashi15}.

\section{Discussion and Conclusions}
\label{sec:disc}

We have analyzed the radiation hydrodynamics of unbound outflows originating from the \ltwo\ point of binary stars, employing realistic equation of state and opacities. The outflowing spiral stream spreads and the spiral arms overlap and merge at a collision radius, where the resulting shock wave thermalizes $\sim 10-20$ per cent of the radial kinetic energy of the outflow (Figs.~\ref{fig:structure} and \ref{fig:velocity}). Depending on the binary properties and mass loss rate, radiative cooling, diffusion, or adiabatic expansion dominate the energy evolution of the outflow (Fig.~\ref{fig:structure}) with effects on the produced luminosity (Fig.~\ref{fig:lum}) and dust formation (Fig.~\ref{fig:transition}). 

We generalize our results to any predominantly equatorial outflow originating from a binary star, where the energy deposition at the inner edge and the asymptotic velocity are both proportional to the binary escape velocity. This is motivated by a some scenarios of the common envelope phase.  We construct a semi-analytic model of such outflows, including luminosity and effective temperature evolution (Sec.~\ref{sec:an}), and investigate the peak luminosities as a function of the timescale (Fig.~\ref{fig:summary}). The produced luminosities (Fig.~\ref{fig:summary}), effective temperatures (Fig.~\ref{fig:summary_teff}), and ejecta velocities (Fig.~\ref{fig:vej}) are able to match red transients such as V838~Mon, V1309~Sco, and OGLE~2002-BLG-360, but we cannot reasonably match the properties of SN2008S, NGC300OT, and PTF10fqs indicating a different physical process for these transients.

Our results are similar to the models based on hydrogen recombination in an expanding shell \citep{ivanova13b,nandez14} in the sense that the pseudophotosphere in our models is often positioned at the greatest opacity gradient corresponding to hydrogen ionization/recombination and that the ejecta velocity is proportional to the binary escape velocity. However, we naturally take into account much wider range of densities and temperatures and implicitly include dust formation. Our model can produce $\teff \lesssim 3000$\,K, which would be hard for a purely hydrogen recombination model, and we can explain the $\sim 850$\,K effective temperature of OGLE~2002-BLG-360 \citep{tylenda13}. Furthermore, we suggest the possibility of bright ($L \sim 10^5\,\lsun$), cool ($\teff \lesssim 1500$\,K), and long ($\tobs \sim 1000$\,days) transients, which would radiate predominantly in the near-IR. Such transients might have been recently discovered with {\em Spitzer} \citep{kasliwal14}.

Although we are able to match the general properties of the observed class of red transients, the evolution of the binary and its orbit is not directly coupled to the outflow in our models. As a result, we cannot reproduce the light curve of individual transients in detail.  For example, V838~Mon showed evidence of multiple mass ejections, possibly interacting with each other \citep{tylenda05}.  Similar complicated behavior is observed in many other red transients. Evolving together the hydrodynamics of binary interactions and radiation hydrodynamics of the resulting outflow is challenging within a single code due to the large range of timescales and physical conditions. Instead, the dynamical results of mass ejection from simulations such as those of \citet{nandez14} could be used as an input for our radiation hydrodynamics model in future work.

Explaining portions of the light curves of some red transients might be feasible even without detailed modeling of the binary hydrodynamics. V1309~Sco exhibited a slow and gentle $\sim 150$\,day rise to the maximum, which can be reconciled with the preceding orbital period evolution by considering the \ltwo\ mass loss, as argued by \citet{pejcha14}. The rate of mass loss from \ltwo\ depends primarily on the physics of the stellar surface layers, and hence a simple power-law runaway model might be sufficient to describe the system and to provide useful constraints on the theory \citep{paczynski72,webbink77,pejcha14}. OGLE~2002-BLG-360 is similar to V1309~Sco in these aspects \citep{tylenda13}. Investigations of these issues with our detailed radiation hydrodynamics models will be the subject of future work.

Another extension of the presented work we are currently pursuing are the optical signatures of the formation of an excretion disk. This should be the outcome of \ltwo\ mass loss for binaries with mass ratios $q \lesssim 0.06$ or $q \gtrsim 0.8$. For these binaries, the outflow remains bound and stalls at a particular radius \citep{shu79}, forming a mass loss ring that continues to interacts with the surface of the binary star. Different inner boundary conditions will be needed to properly investigate the structure of the mass-loss ring and its effect on the central binary. In a preliminary exploration of the early phases with our current inflow boundary condition, we find that such systems can be much more luminous and more efficient in extracting angular momentum from the binary than in the unbound outflow case.  However, a detailed investigation is beyond this scope of this paper and is relegated to future work. Our results might be applicable also to planet-star mergers, which previous work has suggested could give rise to red-nova like transients \citep{metzger12}, and to double white dwarf mergers, which might be the common outcome of their interaction \citep{shen15}.

Finally, we mention the possibility that the shock heating of the ejecta might not be stationary at the inner edge of the outflow, but the shock wave might dynamically propagate through the outflow. This may be a result of a collision of two shells launched with different asymptotic velocity, or due to instantenous deposition of a large amount of energy at the base of the outflow \citep[e.g.][]{sandquist98,morris06}. In both cases, the propagating shock wave will heat the outflow and the resulting electromagnetic emission will depend on the density and velocity profile of the outflow, and the energy deposited. When the shock reaches the photosphere the radiation behind the shock might be suddenly released in a form of a shock breakout. The formalism developed for the shock breakout from an optically-thick wind in the context of interaction-powered supernovae could be applied to this situation \citep[e.g.][]{balberg11,chevalier11,ginzburg12,ginzburg14,svirski12}, however the strong asphericity of the binary outflow will likely alter the transient characteristics.

\section*{acknowledgements}
We thank Chris Kochanek for detailed reading of the manuscript. OP acknowledges discussions with Todd Thompson, Chris Kochanek, Zhaozhuan Zhu, and Andrei Beloborodov. We thank the anonymous referee for constructive comments that helped to improve the paper. Part of the simulations were carried out using computers supported by the Princeton Institute of Computational Science and Engineering. Support for OP was provided by NASA through Hubble Fellowship grant HST-HF-51327.01-A awarded by the Space Telescope Science Institute, which is operated by the Association of Universities for Research in Astronomy, Inc., for NASA, under contract NAS 5-26555. Part of the work of OP was performed at the Aspen Center for Physics, which is supported by National Science Foundation grant PHY-1066293.  BDM acknowledges support from the NSF grant AST-1410950 and the Alfred P. Sloan Foundation.

\appendix
\section{Definition of quantities for radiative diffusion, cooling, and heating}

\subsection{Radiative diffusion}
\label{app:diff}
The radiative diffusion coefficient $k_i$ for particle $i$ is calculated as
\beq
k_i = \frac{16\stefb}{\rho_i\kappa_i}\lambda_i T_i^3,
\label{eq:app_ki}
\eeq
where $\kappa_i$ is the Rosseland-mean opacity and $\lambda_i$ is the flux limiter
\beq
\lambda_i = \frac{2 + R_i}{6 + 3R_i + R_i^2}.
\eeq
The quantity $R_i$ is defined as
\beq
R_i = \frac{|\nabla u_{{\rm r}, i}|}{u_{{\rm r},i}\rho_i\kappa_i},
\eeq
where we calculate the gradient of the radiation energy density $u_{{\rm r}, j} = 4\stefb T_j^4/c$ as
\beq
\nabla u_{{\rm r}, j} = \sum_i \frac{m_i}{\rho_i} u_{{\rm r}, i} \nabla W_{ji}.
\eeq
The calculation of radiative diffusion is very fast, because it depends only on local quantities.

\subsection{Radiative cooling and irradiation by the central star}
\label{app:cool}

\begin{figure*}
\centering
\includegraphics[width=0.47\textwidth]{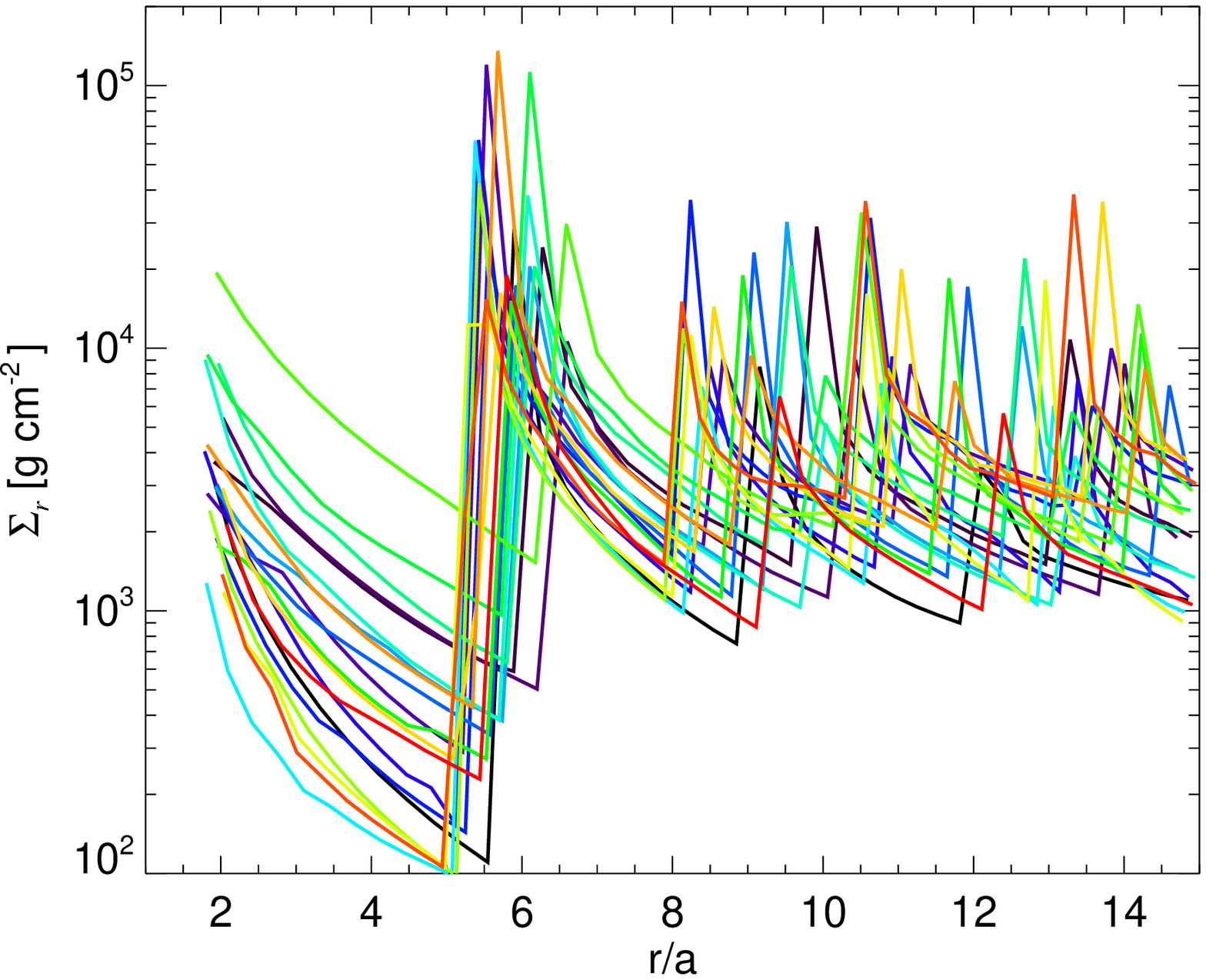}
\includegraphics[width=0.47\textwidth]{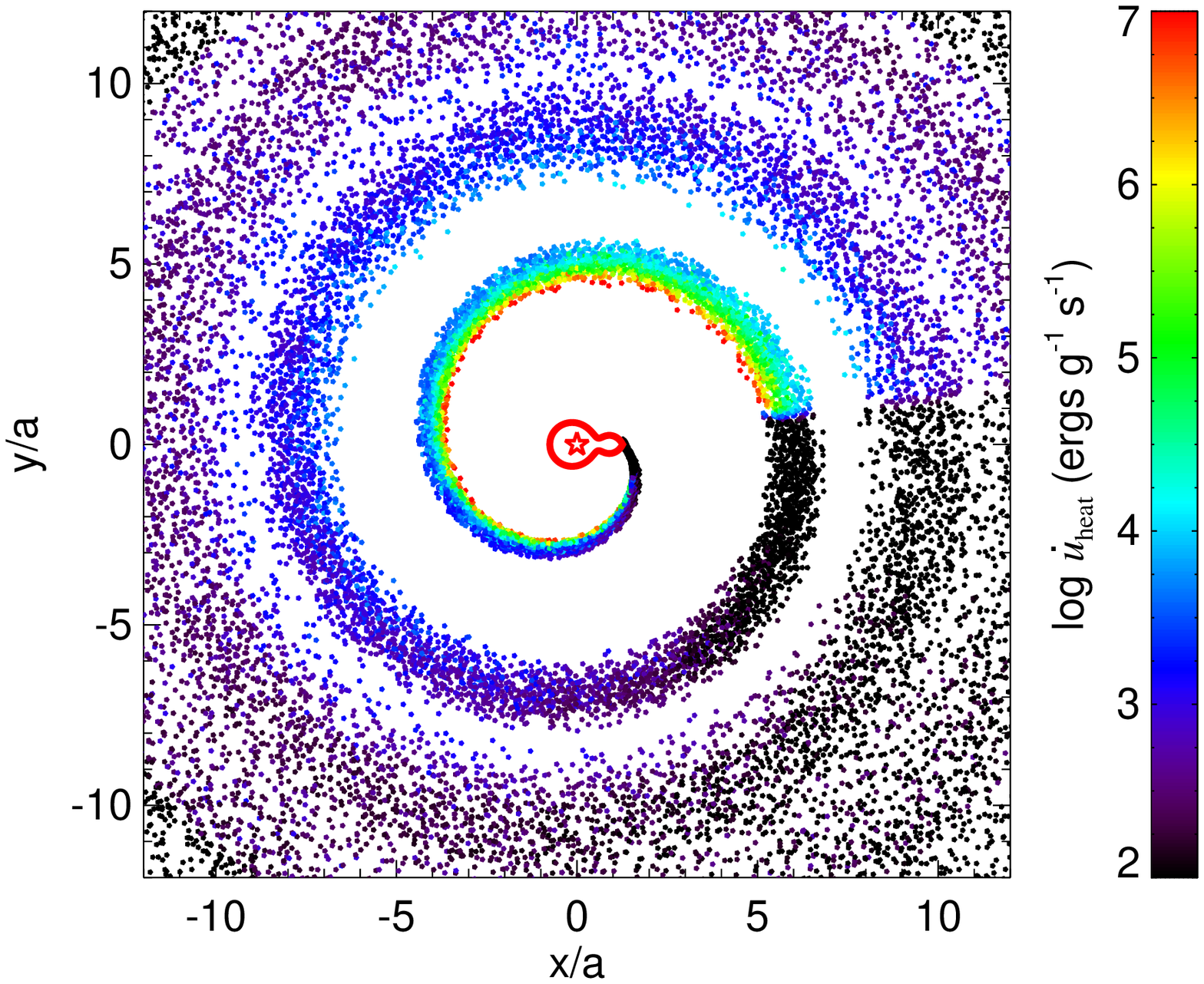}
\caption{Heating rate due to the irradiation from the central star for a calculation with $\mdot = 10^{-3}$\,\myr, $a=0.03$\,AU, $M_2=1.5\,\msun$, $q=0.15$, and $\dot{N} = 3000$. {\em Left} panel shows the radial dependence of $\Sigma_{r,j}$ for $20$ randomly chosen particles. {\em Right} panel shows the heating rate $\dot{u}_{{\rm heat}}$ as a function of position. Each point is a single particle and the color represents $\dot{u}_{\rm heat}$ as indicated by the color bar. The surface of the binary (red closed shape near the center) is replaced by a point source at the barycenter (red star). Following the material as being ejected from $L_2$, the highest $\dot{u}_{\rm heat}$ occurs on the inner rim of the stream and the outer rim is partially self-obscured. The heating rate drops substantially after the spiral starts a second winding and then $\dot{u}_{\rm heat}$ only gradually increases as the interior material expands and becomes more transparent.}
\label{fig:epsdot_heat}
\end{figure*}

We estimate the column density $\Sigma_{z,j}$ and optical depth $\tau_{z,j}$ by integrating $\rho$ and $\kappa\rho$, respectively, along a ray going perpendicularly out of the orbital plane and originating at a particle $j$. Specifically, we use the SPH estimator to evaluate $\rho$ and $\kappa\rho$ along this ray as
\begin{eqnarray}
\Sigma_{z,j} = \int \rho \intd z \approx \sum_{i, z_i/z_j \ge 1} m_i \int_{-\infty}^\infty W(|\bm{r}_j-\bm{r_i}|,h_j)\intd z,\label{eq:sigma_zj}\\
\tau_{z,j} = \int \kappa\rho\intd z \approx \sum _{i, z_i/z_j \ge 1} m_i \kappa_i \int_{-\infty}^\infty W(|\bm{r}_j-\bm{r_i}|,h_j)\intd z,\label{eq:tau_zj}
\end{eqnarray}
where the sums over the particles includes all particles $i$, which are located farther out of the orbital plane than the particle $j$, $z_i/z_j \ge 1$. We do not take into account the contributions to the optical depth from particles with $z_i/z_j <1$, which lie in the smoothing sphere of $j$. We assume that all particles with $z_i/z_j \ge 1$ contribute to the column density and optical depth fully, so that we can integrate over the full smoothing volume. The integral over the smoothing kernel can be expressed analytically, but for the sake brevity we do not reproduce the full expression, and in our calculation, we precompute the values in a lookup table. For the purposes of finding the nearest neighbors of particles along the ray perpendicular to the orbital plane, we project the particles on the orbital plane and construct a two-dimensional tree. This brings a relatively small overhead, because the particles do not get very far from the disk and the hence the number of nearest neighbors is not significantly higher than in three dimensions for the same smoothing length.

To estimate the effects of irradiation of the central star (Eq.~[\ref{eq:heat}]), we calculate the radial column density $\Sigma_{r,j}$ and optical depth $\tau_{r,j}$ as in Equations~(\ref{eq:sigma_zj}--\ref{eq:tau_zj}), with the exception that we are interested in a radial ray. To find the particles along the radial ray, we project all particles on a unit sphere, construct a three-dimensional tree on this projection, and find all nearest neighbors within the smoothing length $h_j/r_j$, effectively obtaining particles in a radial cone. Although the actual construction of the three-dimensional tree is relatively fast, the number of particles in a given cone can be relatively large, because the smoothing length typically increases linearly with the radial distance. As a result, calculating $\dot{u}_{{\rm heat},j}$ is relatively expensive and significantly slows down the calculation. To mitigate this, we include only particles with $r_j \le 15a$. Outside of this radius, the equilibrium temperature of the star is low and the radiative cooling dominates. In Figure~\ref{fig:epsdot_heat}, we show the typical behavior of $\Sigma_{r,j}$ and $\dot{u}_{\rm heat}$.

\section{Dependence on technical parameters}
\label{app:technical}

\begin{figure}
\centering
\includegraphics[width=\columnwidth]{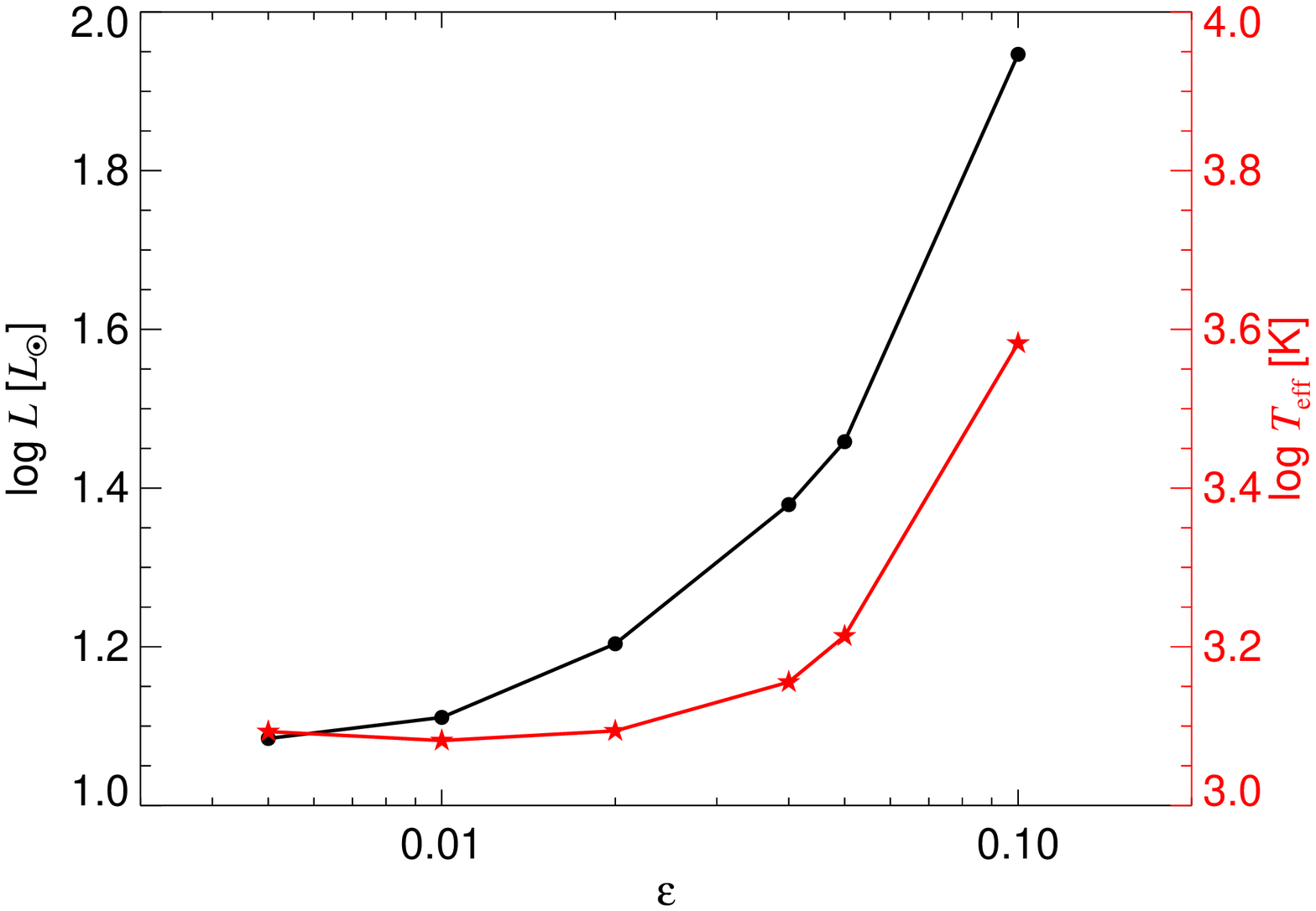}
\caption{Luminosity and effective temperature as a function of the stream spreading parameter $\varepsilon$ for the binary parameters from Figure~\ref{fig:structure}, and $\dot{N} = 3000/P$.\label{fig:epsilon}}
\end{figure}

\begin{figure*}
\centering
\includegraphics[width=0.49\textwidth]{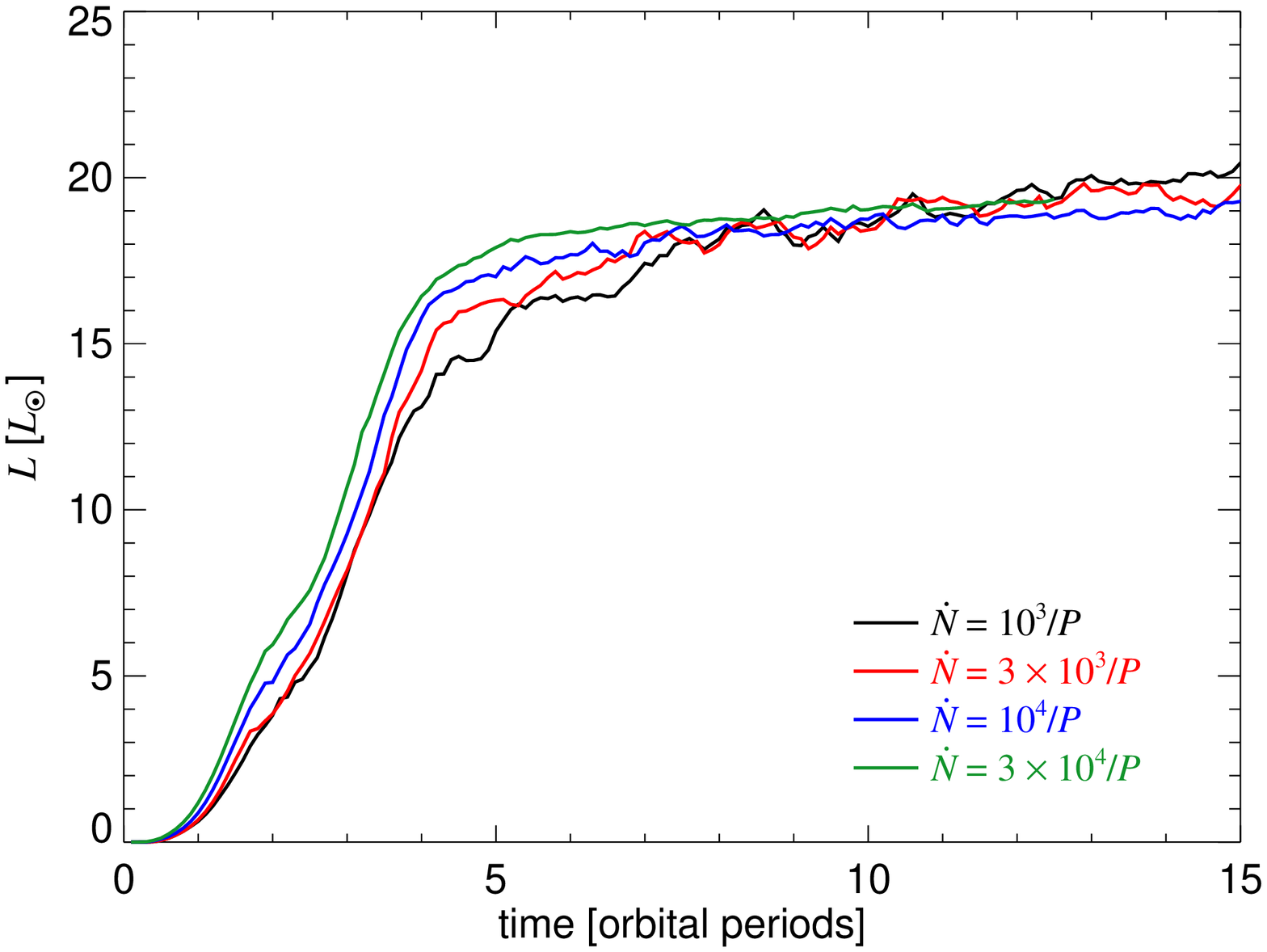}\includegraphics[width=0.49\textwidth]{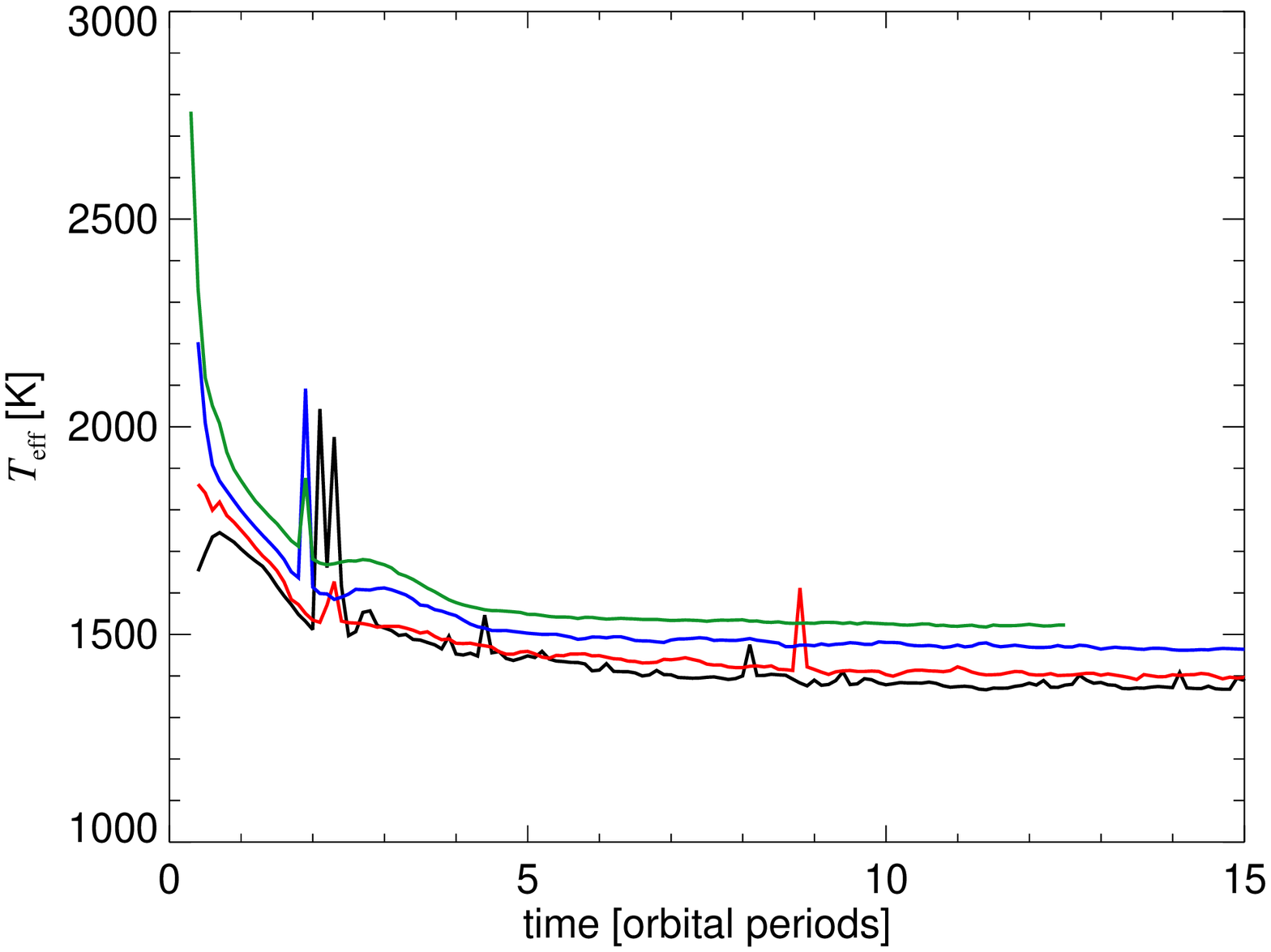}
\caption{Radiative cooling luminosity (left panel) and effective temperature (right panel) as a function of elapsed time for a simulation with $a=0.03$\,AU, $M=0.43\,\msun$, $q=0.15$, and $\mdot=10^{-3}$\,\myr. The lines are for different rates of particle ejection $\dot{N} = 10^3/P$ (black), $3\times 10^3/P$ (red), $10^4/P$ (blue), and $3\times 10^4/P$ (green).}
\label{fig:resolution}
\end{figure*}

In addition to physical parameters, it is worth investigating how our results depend on the technical parameters of the calculation. Although the stream-spreading parameter $\varepsilon$ should be approximately constant for different binary classes (Eq.~[\ref{eq:epsilon}]), the specific value is not known without detailed hydrodynamical study of the binary surface at the \ltwo\ point. In Figure~\ref{fig:epsilon}, we show the dependence of the luminosity and effective temperature on $\varepsilon$. We see that for $\varepsilon \gtrsim 0.02$ both $L$ and $\teff$ increase, because wider stream implies greater velocity spread and higher shock temperatures (Sec.~\ref{sec:an}). For $\varepsilon \lesssim 0.02$ both $L$ and $\teff$ stay approximately constant. The reason is that for too small $\varepsilon$ the particles will be injected in a much smaller region than what would correspond to their temperature, the thermal pressure forces will become important and the stream will expand. Figure~\ref{fig:epsilon} also shows that our fiducial $\varepsilon = 0.05$ perhaps slightly overestimates the appropriate value, but the differences are much smaller than other uncertainties in the input physics and calculation methods.

In Figure~\ref{fig:resolution}, we show the dependence of the luminosity and effective temperature on the rate of particle injection $\dot{N}$, which corresponds to the simulation resolution. The scatter of the lines goes down as $\dot{N}$ increases, especially during the transient epoch when the spiral arms start merging for the first time. Runs with higher $\dot{N}$ also converge to the asymptotic value faster, but the asymptotic luminosities and effective temperatures are independent of resolution. We run all our calculations for at least $15P$, but typically much longer until the steady-state is achieved.

\end{document}

%% file: journal.tex
\def\aj{AJ}%
\def\araa{ARA\&A}%
\def\apj{ApJ}%
\def\apjl{ApJ}%
\def\apjs{ApJS}%
\def\ao{Appl.~Opt.}%
\def\aap{A\&A}%
\def\aapr{A\&A~Rev.}%
\def\mnras{MNRAS}%
\def\pasp{PASP}%
\def\nat{Nature}%
\def\na{New~Astron.}%
\def\physrep{Phys.~Rep.}%
\def\pasa{PASA}
\def\actaa{Acta Astron.}